\documentclass[12pt]{article}
\usepackage{amsmath,amsfonts,amsbsy,amssymb,amsthm,bbm}
\usepackage{multirow}
\usepackage{mathtools}
\usepackage{graphicx,psfrag,epsf}
\usepackage{natbib}
\usepackage{booktabs}  
\usepackage{url}       
\usepackage{enumitem}  
\usepackage{hyperref}
\hypersetup{
  colorlinks,
  citecolor=blue,
  linkcolor=blue,
  urlcolor=blue} 
\usepackage{multibib}  
\usepackage{tikz} 
\usetikzlibrary{backgrounds,positioning,patterns}
\usepackage{xcolor}
\usepackage{comment}
\usepackage{geometry,caption,color,setspace,sectsty,array, physics, mathrsfs, subcaption}
\usepackage{pgfplots} 
\pgfplotsset{compat=1.18}
\usepackage{adjustbox, rotating}
\usepackage{chngcntr}
\usepackage{catchfile}
\usepackage{soul}

\usepackage{etoolbox}
\usepackage[ruled,lined,linesnumbered,longend]{algorithm2e}  
\SetArgSty{textnormal}                
\AtBeginEnvironment{algorithm}{\spacingset{1.2}}  
\SetAlgoSkip{bigskip}                 
\SetAlgoInsideSkip{smallskip}         
\SetInd{0.6em}{1.1em}                 
\SetNlSty{textbf}{}{}                 
\SetNlSkip{0.9em}
\SetAlCapSkip{0.3em}
\DontPrintSemicolon
\SetKwInput{KwIn}{Input}
\SetKwInput{KwOut}{Output}
\SetKwComment{tcp}{$\triangleright$ }{}

\SetCommentSty{algcommentsty}

\newcites{appendix}{References}
\usepackage{todonotes}

\newtheoremstyle{mystyleRM}
  {}
  {}
  {\upshape}
  {}
  {\bfseries}
  {.}
  { }
  {\thmname{#1}\thmnumber{ #2}\thmnote{ (#3)}}

\theoremstyle{mystyleRM} 

\newtheorem{theorem}{Theorem}[section]
\newtheorem{proposition}{Proposition}[section]

\newtheorem{lemma}{Lemma}[section]

\newtheorem{definition}{Definition}[section]

\newtheorem{remark}{Remark}
\newtheorem{example}{Example}

\newcommand{\blind}{0}

\graphicspath{{./Figures/}{../R_Codes/Applications/Figures/}}

\newcommand{\maybeinclude}[2][]{\IfFileExists{./Figures/#2}{\includegraphics[#1]{#2}}{\IfFileExists{../R_Codes/Applications/Figures/#2}{\includegraphics[#1]{#2}}{\fbox{\parbox[c][0.25\textheight][c]{0.9\linewidth}{\centering\small Figure pending:\\ \texttt{\detokenize{#2}}\\ (run \texttt{Application.R})}}}}}

\newcolumntype{C}[1]{>{\centering\arraybackslash}p{#1}}

\newcommand\underlay[4]{%
  \stackengine{0pt}%
  {\kern#2\includegraphics[height=#1]{#4}}%
  {\includegraphics[height=#1]{#3}}%
  {O}{l}{F}{F}{L}%
}
\newcommand\addunderlay[4]{%
  \stackengine{0pt}%
  {\kern#2\includegraphics[height=#1]{#4}}%
  {#3}%
  {O}{l}{F}{F}{L}%
}

\newcommand*{\colorboxed}{}
\def\colorboxed#1#{%
  \colorboxedAux{#1}%
}
\newcommand*{\colorboxedAux}[3]{%
  \begingroup
    \colorlet{cb@saved}{.}%
    \color#1{#2}%
    \boxed{%
      \color{cb@saved}%
      #3%
    }%
  \endgroup
}

\renewcommand{\arraystretch}{1.5}

\NewCommandCopy{\proofqedsymbol}{\qedsymbol}

\AtBeginEnvironment{proof}{\renewcommand{\qedsymbol}{\proofqedsymbol}}

\theoremstyle{definition}

\AtBeginEnvironment{example}{%
  \pushQED{\qed}\renewcommand{\qedsymbol}{$\blacksquare$}
}
\AtEndEnvironment{example}{\popQED\endexample}
\date{}

\begin{document}

\def\spacingset#1{\renewcommand{\baselinestretch}%
{#1}\small\normalsize} \spacingset{1}


\if0\blind
{
  \title{\bf Anytime Valid Conformal Prediction on Streaming Time Series}
  \author{
  Chencheng Fang\thanks{Institute of Finance and Statistics, University of Bonn, Adenauerallee 24-42, 53113 Bonn, Germany}
  \thanks{Hausdorff Center for Mathematics, Endenicher Allee 62, 53115 Bonn, Germany}
  }
  \maketitle
} \fi

\if1\blind
{
  \bigskip
  \bigskip
  \bigskip
  \begin{center}
    {\LARGE\bf Anytime Valid Conformal Prediction on Streaming Time Series}
\end{center}
  \medskip
} \fi

\begin{abstract}
  Sequential prediction on streaming time series is essential for continuous monitoring and risk control in applications such as finance, energy, and healthcare systems, where prediction intervals must remain valid at arbitrary stopping times and adapt to potential distributional shifts. Conventional conformal prediction provides model-agnostic uncertainty quantification but relies on exchangeability, which is violated under distributional shifts. Adaptive Conformal Inference (ACI) relaxes this assumption by updating the miscoverage level in response to the realized prediction errors, but it only guarantees that the realized average miscoverage rate along a single trajectory approaches the target level in the long run. Risk-sensitive monitoring instead requires controlling the true miscoverage rate, i.e., the conditional probability that a prediction set misses, which measures the reliability of the set itself rather than the misses that happened to be realized. We show that, without assumptions on the distributional shifts, it is impossible to have such control for every single prediction set at arbitrary stopping times. We therefore target the true running and rolling average miscoverage rates. Our methods combine ACI with anytime valid confidence sequences. With high probability, the true running average miscoverage rate stays below the target level uniformly over time, and so does the true rolling average for any window rule. The efficiency loss for the rolling average vanishes for windows that grow sufficiently fast, but not for fixed windows.
\end{abstract}


\spacingset{1.5}

\section{Introduction}\label{sec:intro}

A streaming time series refers to an ordered and unlimited sequence of observations that arrive one at a time, rather than in batches. It arises naturally in settings where decisions must be made in real time. For example, electrical grid operations require continuous monitoring, real-time coordination, and instant decision-making to ensure the reliability of power dispatch. Cardiac surveillance involves continuously tracking heart rhythms, detecting abnormal heartbeats, and calling for prompt medical intervention in potentially serious conditions. In such settings, a practitioner usually does not wait for a fixed sample to accumulate before acting. Instead, decisions are triggered by the streaming data as it arrives, at time points that are themselves data-dependent. For a recent review on streaming time series, see \citet{Melgar_2026}. 

Sequential prediction is a fundamental topic on streaming time series, aiming to continuously predict future responses from the sequence of historical data. Traditional time series prediction methods, which rely on fixed datasets and offline training, may be ill-suited to streaming environments that require continuous model adaptation, timely detection of drifts, and effective modeling of temporal dependencies. Machine Learning (ML) methods, however, have demonstrated strong predictive performance in streaming environments by continuously updating model parameters as new data arrive, adapting to drifts, and capturing temporal dependencies \citep{gomes_2019, Melgar_2026}. Despite the predictive flexibility, many ML methods typically lack inherent mechanisms for constructing valid prediction intervals.

Conformal prediction offers a principled, distribution-free framework for constructing prediction sets for future responses with a marginal finite-sample validity guarantee \citep{papadopoulos_2002, vovk_gammerman_shafer_2005, lei_2018, angelopoulos_bates_2023}. Crucially, it is model-agnostic: the validity guarantee holds regardless of the prediction method, making it potentially well-suited for uncertainty quantification of ML prediction methods used in streaming environments. Split conformal prediction is one of the most popular and computationally tractable conformal prediction algorithms (see Section \ref{ssec:full_split_cp}). It generates a prediction set $\mathcal{C}_{n}(X)$ using a held-out calibration set $\mathcal{D}_{n_2}$, which is a part of the whole training set $\mathcal{D}_n=(X_i,Y_i)_{i=1}^n$, and guarantees that the miscoverage rate is no greater than a target level $\alpha$ for the next observation $(X_{n+1}, Y_{n+1})$, i.e., 
\begin{equation}\label{eq:cp_validity}
  \mathbb{P}(Y_{n+1} \not\in \mathcal{C}_{n}(X_{n+1})) \le \alpha.
\end{equation}
A widely discussed limitation of split conformal prediction is that the validity guarantee is merely a marginal control \textit{jointly} over the calibration set $\mathcal{D}_{n_2}$ and the test point $(X_{n+1}, Y_{n+1})$, i.e.,
\begin{equation*}
  \mathbb{E}[\mathbb{P}( Y_{n+1} \not\in \mathcal{C}_{n}(X_{n+1}) \mid \mathcal{D}_{n_2})]=\mathbb{P}(Y_{n+1} \not\in \mathcal{C}_{n}(X_{n+1})) \le \alpha.
\end{equation*}
The marginal guarantee is an average over repeated draws of the calibration set from the same distribution. However, in practice, when there is only one single realized calibration set, the conditional miscoverage rate can deviate substantially above $\alpha$ \citep{vovk_2012, marques_2025, hultberg_2026}. This motivates a large body of research on a more appealing and stronger validity guarantee on the calibration-conditional miscoverage rate. 

\paragraph{Related Literature.}
The first branch of literature on the calibration-conditional miscoverage rate relates to classical tolerance regions \citep{krishnamoorthy_2008} and the so-called Probably Approximately Correct (PAC) prediction sets. The goal is to generate a prediction set $\mathcal{C}_{n}$ such that the calibration-conditional miscoverage rate is approximately correct with high probability, i.e., for some small $\delta>0$,
\begin{equation}\label{eq:pac}
  \mathbb{P}(\mathbb{P}( Y_{n+1} \not\in \mathcal{C}_{n}(X_{n+1}) \mid \mathcal{D}_{n_2}) \le \alpha) \ge 1-\delta.
\end{equation}
\cite{vovk_gammerman_shafer_2005} show that the worst-case calibration-conditional miscoverage probability over all data-generating processes equals one unless the prediction set is the whole real line, and hence the calibration-conditional guarantee cannot hold with certainty and should take the PAC form as in \eqref{eq:pac}. Many studies assume the calibration set $\mathcal{D}_{n_2}$ and the test point $(X_{n+1}, Y_{n+1})$ to be independent and identically distributed (IID), and rely on classical concentration inequalities to derive the prediction set $\mathcal{C}_{n}$ that satisfies \eqref{eq:pac} \citep{vovk_2012,bates_2021,duchi_2025}. Some studies relax the IID assumption and derive such a prediction set under covariate shifts by reweighting the calibration data with a likelihood ratio between the test and calibration covariate distributions, which is assumed known \citep{park_2022, pournaderi_xiang_2026}. \cite{qiu_2023} estimate the likelihood ratio from data, but their guarantee then holds only asymptotically. All of these results, however, concern the prediction set at a single time, making them unsuitable for sequential prediction on streaming time series.

The second branch of literature on the calibration-conditional miscoverage rate is thus concerned with a validity guarantee over an ordered and unlimited sequence of prediction sets. At any stopping time $\tau\ge1$, a prediction set $\mathcal{C}_{\tau-1}$ for the response $Y_\tau$ is constructed from the observations $\mathcal{D}_{\tau-1}=\{(X_1,Y_1),\dots,(X_{\tau-1},Y_{\tau-1})\}$ up to time $\tau-1$. The goal is to sequentially generate a series of prediction sets $\{\mathcal{C}_{\tau-1}\}_{\tau\ge1}$ such that the calibration-conditional miscoverage rates of all prediction sets at arbitrary stopping times are controlled uniformly with high probability, i.e., for some small $\delta>0$,
\begin{equation}\label{eq:pac_anytime}
  \mathbb{P}( \forall \tau \ge 1: \mathbb{P}( Y_{\tau} \not\in \mathcal{C}_{\tau-1}(X_{\tau}) \mid \mathcal{D}_{\tau-1}) \le \alpha) \ge 1-\delta.
\end{equation}
Similarly, the derivation of such prediction sets in many studies depends on the IID assumption. For example, \cite{avelin_2023} develops sequential inductive prediction intervals based on a time-uniform version of Dvoretzky-Kiefer-Wolfowitz inequality from \cite{howard_2022}. To extend both the conformal prediction and PAC frameworks to the sequential setting, \cite{scharfstein_2026} construct the prediction set at each time with a more conservative empirical quantile, chosen either from a confidence sequence for a fixed quantile from \cite{howard_2022}, or from a union bound over time. \cite{xu_karampatziakis_mineiro_2024} adapt risk-controlling prediction sets to the sequential scenario via testing by betting, a game-theoretic framework advocated by \cite{shafer_2021} in which the evidence against a hypothesis is measured by the accumulated wealth of a gambler who repeatedly bets against it. Some recent works relax the IID assumption and discuss deriving the prediction sets under distributional shifts. For instance, \cite{hultberg_2026} develop the anytime valid conformal risk control that accommodates distributional shifts by reweighting each calibration observation with a known importance weight, namely the likelihood ratio between the test and calibration distributions. In practice, however, the importance weight is rarely known, and estimating it introduces errors that invalidate the finite-sample guarantee. 

Therefore, a question arises as to whether we can generate a sequence of prediction sets that satisfies \eqref{eq:pac_anytime} under distributional shifts without any prior knowledge of the shifts. The answer is negative. In Proposition \ref{prop:no.pac}, we show that, without any restrictive assumption on the nature of distributions, it is impossible to derive such prediction sets. Since a prediction set for the stopping time $\tau$ is constructed before the true response at time $\tau$ is realized, an unanticipated distributional shift at time $\tau$ can move the true response so far from the prediction set that the set misses with probability close to one. A single such failure, however, has little effect on the \emph{average} of calibration-conditional miscoverage rates over the whole period up to time $\tau$. \cite{xu_karampatziakis_mineiro_2024} also conjecture that, under distributional shifts, the risk can be controlled ``in an average sense''. 

\paragraph{Targets and Inferential Goals.}
In this paper, we target two such averages: a \emph{running} average over all periods up to any stopping time $\tau$ from the start of monitoring, and a \emph{rolling} average over the most recent $k_\tau$ periods up to time $\tau$, where the window length $k_\tau$ may depend on $\tau$. The goal is to sequentially generate prediction sets $\{\mathcal{C}_{\tau-1}\}_{\tau\ge1}$ such that the running or rolling averages of the calibration-conditional miscoverage rates up to any stopping time $\tau\ge1$ are controlled uniformly with high probability, i.e., for some small $\delta>0$,
\begin{alignat}{2}
  &\text{(Running)}\quad && \mathbb{P}\left(\forall\tau\ge1:\ \frac{1}{\tau}\sum_{t=1}^{\tau}\nu_t \le\alpha\right)\ge1-\delta,\label{eq:running}\\
  &\text{(Rolling)}\quad && \mathbb{P}\left(\forall\tau\ge1:\ \frac{1}{k_\tau}\sum_{t=\tau-k_\tau+1}^{\tau}\nu_t \le\alpha\right)\ge1-\delta,\label{eq:rolling}
\end{alignat}
where $\nu_t=\mathbb{P}( Y_t \not\in \mathcal{C}_{t-1}(X_t) \mid \mathcal{D}_{t-1})$ and $\tau \ge k_\tau$. We refer to $\nu_t$ as the true miscoverage rate of the prediction set at time $t$. The validity guarantee in \eqref{eq:running} describes the reliability of a procedure over its whole lifetime, while the guarantee in \eqref{eq:rolling} measures its recent reliability. The two can differ substantially: a single coverage failure barely moves a running average over several months, but it dominates a rolling average over a week. To construct prediction sets that satisfy \eqref{eq:running} and \eqref{eq:rolling}, our methods build on two theoretical tools in the literature.

\paragraph{Methods.}
The first tool we rely on is Adaptive Conformal Inference \citep[ACI,][]{gibbs_candes_2021}. The validity guarantee of split conformal prediction, as shown in \eqref{eq:cp_validity}, depends on an exchangeability assumption, which states that the joint distribution of observations is invariant to permutations (see Definition \ref{def:ex}). This assumption is routinely violated under temporal dependency and distributional shifts. ACI is a pioneering approach to addressing the loss of validity of split conformal prediction under distributional shifts (see Section \ref{ssec:aci}). Owing to its theoretical simplicity and computational tractability, ACI has inspired many subsequent methods, including Online Expert Aggregation on ACI \citep[AgACI,][]{zaffran_2022}, Dynamically-tuned Adaptive Conformal Inference \citep[DtACI,][]{gibbs_candes_2024}, and Online Conformal Inference with Retrospective Adjustment \citep[RetroAdj,][]{jun_2026}. It handles the challenge by updating the miscoverage level in response to the observed coverage errors, i.e., set $\alpha_1=\alpha$, for $t \ge1$,
\begin{equation}\label{eq:intro.aci}
  \alpha_{t+1}=\alpha_{t}+\gamma (\alpha-Z_t),
\end{equation}
where $Z_t=\mathbbm{1}\{Y_t \not\in \mathcal{C}_{t-1,\alpha_t}(X_{t})\}$ is an indicator equal to one if the prediction set misses the true response at time $t$, and $\gamma>0$ is a learning rate. The subscript $\alpha_t$ in $\mathcal{C}_{t-1,\alpha_t}$ indicates that the prediction set is computed with the miscoverage level $\alpha_t$. As implied by \cite{gibbs_candes_2021}, without any assumption on the distributional shifts, ACI always ensures that the \emph{realized} average miscoverage rate exceeds the target level $\alpha$ by at most a vanishing term for any $\tau\ge1$:
\begin{equation}\label{eq:intro.aci.realized}
    \frac{1}{\tau}\sum_{t=1}^{\tau}Z_t \le \alpha + \frac{\alpha + \gamma}{\gamma \tau}.
\end{equation}
In particular, the realized average miscoverage rate is asymptotically at most $\alpha$. The original ACI, however, cannot achieve the validity guarantee in \eqref{eq:running} for two reasons. First, the upper bound in \eqref{eq:intro.aci.realized} contains an excess of $(\alpha+\gamma)/(\gamma\tau)$ over $\alpha$, which takes into account the misses ACI makes before it adapts to a shift. Second, the guarantee in \eqref{eq:intro.aci.realized} concerns the realized average miscoverage rate, not the true average miscoverage rate. Given $\mathcal{D}_{t-1}$, each $Z_t$ is actually a single draw of the miss with probability $\nu_t$, and hence the true average can exceed the realized one by a statistical error that must be controlled uniformly over time. 

The second tool we rely on is the anytime valid confidence sequence, which is a sequence of confidence sets that is uniformly valid over an unbounded time horizon. The foundations date back to \cite{darling_1967}, who develop iterated logarithm inequalities for sequential inference. This line of work has been refined and extended, especially in recent years. For example, \cite{howard_2020} develop a general framework of time-uniform Chernoff bounds based on nonnegative supermartingales, which covers a range of concentration inequalities. \cite{howard_2021} combine \cite{howard_2020}'s linear boundaries into curved ones by stitching and mixing, which yields closed-form confidence sequences that remain anytime valid even when the conditional means change over time. \cite{waudby_2024} derive time-uniform confidence sequences for estimating means of bounded random variables by betting. \cite{chugg_ramdas_2025} propose closed-form empirical Bernstein confidence sequences for scalars and matrices, the width of which adapts to the observed variability. 

In this paper, since $\mathbb{E}[Z_t \mid \mathcal{D}_{t-1}]=\nu_t$, the differences $\{\nu_t-Z_t\}_{t\ge1}$ actually form a bounded martingale difference sequence, regardless of distributional shifts. It allows us to derive an anytime valid confidence sequence for the true average miscoverage rate from the realized average miscoverage rate, i.e., for a small $\delta>0$,
\begin{equation}\label{eq:intro.cs}
  \mathbb{P}\left(\forall\tau\ge1:\ \frac{1}{\tau}\sum_{t=1}^{\tau}\nu_t\le\frac{1}{\tau}\sum_{t=1}^{\tau}Z_t+\omega_\tau\right)\ge1-\delta,
\end{equation}
where $\omega_\tau$ is the margin of the confidence sequence, which vanishes as $\tau\to\infty$. For the original ACI, combining \eqref{eq:intro.aci.realized} and \eqref{eq:intro.cs} gives
\begin{equation}\label{eq:intro.aci.cs}
  \mathbb{P}\left(\forall\tau\ge1:\ \frac{1}{\tau}\sum_{t=1}^{\tau}\nu_t\le\alpha+\frac{\alpha+\gamma}{\gamma\tau}+\omega_\tau\right)\ge1-\delta .
\end{equation}
The upper bound in \eqref{eq:intro.aci.cs} exceeds $\alpha$ by two terms: the excess $(\alpha+\gamma)/(\gamma\tau)$ of ACI and the margin $\omega_\tau$ of the confidence sequence.

To remove both additional terms in \eqref{eq:intro.aci.cs} as compared to our goal in \eqref{eq:running}, we propose a new updating rule for the miscoverage level of ACI. Summing up both sides of the original updating rule in \eqref{eq:intro.aci} over time $1$ to $t$, and rearranging the terms yields 
\begin{equation}\label{eq:sum.aci}
  \alpha_{t+1}=\alpha+\gamma(\alpha t-S_t),
\end{equation}
where $S_t=\sum_{s=1}^{t}Z_s$. As shown in \eqref{eq:sum.aci}, the update of the miscoverage level in the original ACI is driven by the \textit{realized} sum of misses $S_t$ alone: if the realized sum $S_t$ exceeds the allowance $\alpha t$, the next miscoverage level is smaller than $\alpha$, leading to a wider prediction set and fewer misses afterwards. Otherwise, the next miscoverage level is greater than $\alpha$. ACI thus steers $S_t$ towards $\alpha t$. The two additional terms in \eqref{eq:intro.aci.cs} arise because $S_t$ may overshoot $\alpha t$ by $(\alpha+\gamma)/\gamma$ before ACI adapts, and because the scaled margin $t\omega_t$ comes on top of $S_t$ to attain anytime validity. To address this issue, our new updating rule uses a \textit{certified} sum of misses $S^c_t=S_t+t\omega_t$ and a buffer $B$, i.e., 
\begin{equation}\label{eq:sum.aci.certified}
  \eta_{t+1}=\alpha+\gamma(\alpha t-S^c_t-B),
\end{equation}
where $B \ge \alpha/\gamma-\alpha+c$ is a buffer reserved for the overshoot of ACI before it adapts, and $c \ge 1$ bounds the one-step increase of the certified sum of misses such that $S^c_t-S^c_{t-1}\le c$ and depends only on the choice of the anytime valid confidence sequence. The buffer $B>0$ by definition. In Section \ref{ssec:running}, we show that the new updating rule in \eqref{eq:sum.aci.certified} leads to our validity guarantee in \eqref{eq:running} for the true running average miscoverage rate. The construction of prediction sets that satisfies the validity guarantee in \eqref{eq:rolling} for the rolling average follows the same lines and is given in Section \ref{ssec:rolling}. 

\paragraph{Paper Structure.}
The paper and its contributions are organized as follows. Section \ref{sec:cp} reviews split conformal prediction and ACI. In Sections \ref{sec:setup} and \ref{sec:theory}, we present our core theoretical and methodological contributions: Proposition \ref{prop:no.pac} shows that it is impossible to derive bounded prediction sets that satisfy the per-step guarantee in \eqref{eq:pac_anytime} under potential distributional shifts without additional assumptions on their nature. We therefore aim to achieve anytime validity for the true running and rolling average miscoverage rates in \eqref{eq:running} and \eqref{eq:rolling} instead. Lemma \ref{lem:skip} provides anytime valid confidence sequences that bound these true averages by the realized averages plus a margin. Building upon these results, Theorem \ref{thm:run} shows that Algorithm \ref{alg:run} controls the true running average miscoverage rate under the target level at any stopping time. Algorithm \ref{alg:roll} adapts the construction to rolling windows, and Theorem \ref{thm:roll} shows that it controls the true rolling average miscoverage rate, although long windows are needed for the prediction sets to be informative. Simulations in Section \ref{sec:simulations} exhibit the excellent anytime valid properties of our prediction sets for streaming time series with potential distributional shifts. Section \ref{sec:discussion} concludes this paper.

\section{Warmup: Conformal Prediction}\label{sec:cp}

\subsection{Full and Split Conformal Prediction}\label{ssec:full_split_cp}

Conformal prediction \citep{papadopoulos_2002, vovk_gammerman_shafer_2005} offers a general and powerful framework for finite-sample distribution-free uncertainty quantification for a response $Y\in\mathcal{Y}$ given features $X\in\mathcal{X}$. For concreteness, one may take the response space to be $\mathcal{Y}=\mathbb{R}$ and the feature space to be $d$-dimensional $\mathcal{X}=\mathbb{R}^d$. Given a training dataset $\mathcal{D}_n=\{(X_i,Y_i)\}_{i=1}^n$ and a new feature $X_{n+1}$, the goal is to construct a prediction set $\mathcal{C}_{n}(X_{n+1})$ for the new response $Y_{n+1}$, whose miscoverage probability does not exceed the target level $\alpha$ for any sample size $n$. This validity guarantee does not rely on large-$n$ approximations.

Full conformal prediction is the classical approach proposed by \cite{vovk_gammerman_shafer_2005}. It constructs the prediction set by testing, for every \textit{candidate} value $y\in\mathcal{Y}$ of the yet unknown response $Y_{n+1}$, whether $y$ should be included in the set. For each candidate $y$, we first append the hypothetical data point $(X_{n+1},y)$ to the training dataset $\mathcal{D}_n$ and fit a prediction model $\hat{f}_{n,(X_{n+1},y)}: \mathcal{X}\to \mathcal{Y}$ on the augmented dataset $\mathcal{D}_n\cup\{(X_{n+1},y)\}$. Second, we compute nonconformity scores $R:\mathcal{X}\times\mathcal{Y}\to\mathbb{R}$ for all $n+1$ data points. The nonconformity score measures how unusual a data point is relative to the fitted prediction model, with larger values indicating poorer agreement with the model. A common choice of scores is the absolute residual, i.e.,
\begin{align*}
  \begin{aligned}
    R_i &= |Y_i - \hat{f}_{n,(X_{n+1},y)}(X_i)|, \quad i=1,\dots,n\\
    R_{n+1} &= |y-\hat{f}_{n,(X_{n+1},y)}(X_{n+1})|.
  \end{aligned}
\end{align*}
Other examples of scores include scaled residuals, one-sided residuals, and bootstrap-based scores. See Sections 2.2--2.4 and 4.3 in \cite{vovk_gammerman_shafer_2005} for details. Third, after running the first two steps for all candidate values, we define the full conformal prediction set for the test point as
\begin{equation}\label{eq:full.conf.pred.set}
  \mathcal{C}^{\mathrm{full}}_{n}(X_{n+1})=\left\{y \in \mathcal{Y}: R_{n+1} \le \text{the } \lceil(1-\alpha)(n+1)\rceil \text{-th smallest of } R_1,\dots,R_n \right\}.
\end{equation}
The full conformal prediction set $\mathcal{C}^{\mathrm{full}}_{n}(X_{n+1})$ in \eqref{eq:full.conf.pred.set} satisfies the marginal validity guarantee 
\begin{equation}\label{eq:marginal_validity}
  \mathbb{P}(Y_{n+1} \not\in \mathcal{C}^{\mathrm{full}}_{n}(X_{n+1})) \le \alpha
\end{equation}
with only one assumption: exchangeability. Exchangeability is weaker than IID. However, full conformal prediction is computationally very expensive, as the prediction model must be refitted for every candidate value $y\in\mathcal{Y}$.

\begin{definition}[Exchangeability]\label{def:ex}
  Random variables $W_1,\dots,W_m$ are exchangeable if their joint distribution is invariant under permutations, i.e., $(W_1,\dots,W_m)\stackrel{d}{=}(W_{\sigma(1)},\dots,W_{\sigma(m)})$ for any permutation $\sigma$.
\end{definition}

Split conformal prediction \citep{papadopoulos_2002, lei_2018} is a much simpler and more computationally tractable algorithm than full conformal prediction. First, we split the training dataset $\mathcal{D}_n$ into two subsets: $\mathcal{D}_{n_1}$, known as a \textit{proper training set} and $\mathcal{D}_{n_2}$, known as a \textit{calibration set}, such that $\mathcal{D}_{n_1}\cap\mathcal{D}_{n_2}=\emptyset$ and $\mathcal{D}_{n_1}\cup\mathcal{D}_{n_2}=\mathcal{D}_{n}$. Let $n_1=|\mathcal{D}_{n_1}|$ and $n_2=|\mathcal{D}_{n_2}|$. We fit a prediction model $\hat{f}_{n_1}$ on the proper training set $\mathcal{D}_{n_1}$ only once, and use the fitted model to predict on all $n_2$ data points in the calibration set $\mathcal{D}_{n_2}$. Second, we compute nonconformity scores $R:\mathcal{X}\times\mathcal{Y}\to\mathbb{R}$ on the calibration set $\mathcal{D}_{n_2}$, for instance, using the absolute residual:
\begin{equation}\label{eq:split.score}
  R_i=|Y_i-\hat{f}_{n_1}(X_i)|, \quad i\in\mathcal{D}_{n_2}.
\end{equation}
Third, we define the fitted quantile of the nonconformity scores:
\begin{equation*}
  \widehat{Q}(p):=\inf \left\{ r\in \mathbb{R}: \left( \frac{1}{n_2 } \sum_{(x,y)\in \mathcal{D}_{n_2}} \mathbbm{1}_{\{R(x,y) \le r\}}\right)\ge p \right\},
\end{equation*}
and the split conformal prediction set $\mathcal{C}^{\mathrm{split}}_{n}(X_{n+1})$ for the test point:
\begin{equation}\label{eq:split.conf.pred.set}
  \mathcal{C}^{\mathrm{split}}_{n}(X_{n+1})=\left\{y \in \mathcal{Y}: R(X_{n+1},y) \le \widehat{Q} \left( (1-\alpha)(1+\frac{1}{n_2}) \right) \right\},
\end{equation}
where the factor, $1+1/n_2$, is a finite-sample correction that accounts for the unknown score of the test point. If the absolute residual in \eqref{eq:split.score} is used to compute scores, the prediction set \eqref{eq:split.conf.pred.set} takes a simpler form as
\begin{equation}\label{eq:pred.set.absolute.resid}
  \mathcal{C}^{\mathrm{split}}_{n}(X_{n+1})=\left [\hat f_{n_1}(X_{n+1}) \pm \widehat Q\left((1-\alpha)(1+\frac{1}{n_2})\right)\right ].
\end{equation}
Under the exchangeability assumption, the split conformal prediction set $\mathcal{C}^{\mathrm{split}}_{n}(X_{n+1})$ also satisfies the marginal validity guarantee 
\begin{equation}\label{eq:marginal_validity_split}
  \mathbb{P}(Y_{n+1} \not\in \mathcal{C}^{\mathrm{split}}_{n}(X_{n+1})) \le \alpha.
\end{equation}

Split conformal prediction is less computationally expensive than full conformal prediction, because the model is fitted only once, rather than refitted for every candidate value of the response variable. However, split conformal prediction fits the model only on the proper training set, a subset of the whole training dataset, which may reduce statistical efficiency. However, when the training dataset is large, this loss can be outweighed by the gain in computational tractability.

\subsection{Adaptive Conformal Inference}\label{ssec:aci}

The marginal validity guarantees of both full and split conformal prediction rely on the exchangeability assumption, which may not hold in time series problems due to possible temporal dependence and distributional shifts. Existing literature has discussed the extension of conformal prediction theory to time series scenarios. Adaptive Conformal Inference (ACI) is a pioneering approach to account for the setting where the data-generating distribution changes over time in an unknown fashion, and has given rise to a large body of subsequent works, including this paper.

\begin{figure}[!htb]
  \centering
  \begin{tikzpicture}[x=1cm,y=1cm,line join=round,
      bar/.style={draw=black!80,rounded corners=1.5pt,line width=0.5pt},
      unused/.style={bar,fill=white},
      train/.style={bar,fill=orange!25,postaction={pattern=north east lines,pattern color=orange!80!black}},
      calib/.style={bar,fill=cyan!15,postaction={pattern=north west lines,pattern color=cyan!60!blue}},
      test/.style={bar,fill=violet!45}]
    \def\L{3}      
    \def\h{0.42}   
    \def\d{0.5}    
    \foreach \y/\k in {0/0, -0.6/1, -1.7/4, -2.3/5}{
      \ifdim\k pt>0pt \draw[unused] (0,\y) rectangle ++(\k*\d,\h); \fi
      \draw[train] (\k*\d,\y) rectangle ++(\L,\h);
      \draw[calib] (\k*\d+\L,\y) rectangle ++(\L,\h);
      \draw[test] (\k*\d+2*\L,\y) rectangle ++(\d,\h);
    }
    \draw[dotted,thick] (3.6,-0.75) -- (4.5,-1.15);
    \draw[|->] (0,-2.65) -- (9.4,-2.65);
    \foreach \x/\lab in {0/{-T+1},6/{0},9/{t}}
      \draw (\x,-2.55) -- (\x,-2.75) node[below] {$\lab$};
    \node[unused,minimum size=0.3cm,inner sep=0pt] (l1) at (0.35,0.9) {};
    \node[right=1pt of l1] (t1) {\small Unused data};
    \node[train,minimum size=0.3cm,inner sep=0pt,right=8pt of t1] (l2) {};
    \node[right=1pt of l2] (t2) {\small $\mathcal{D}_{T_1}$};
    \node[calib,minimum size=0.3cm,inner sep=0pt,right=8pt of t2] (l3) {};
    \node[right=1pt of l3] (t3) {\small $\mathcal{D}_{T_2}$};
    \node[test,minimum size=0.3cm,inner sep=0pt,right=8pt of t3] (l4) {};
    \node[right=1pt of l4] {\small Test data point};
  \end{tikzpicture}
  \caption{Scheme of OSSCP, adapted from Figure 5(a) in \citet{zaffran_2022}. }
  \label{fig:osscp}
\end{figure}
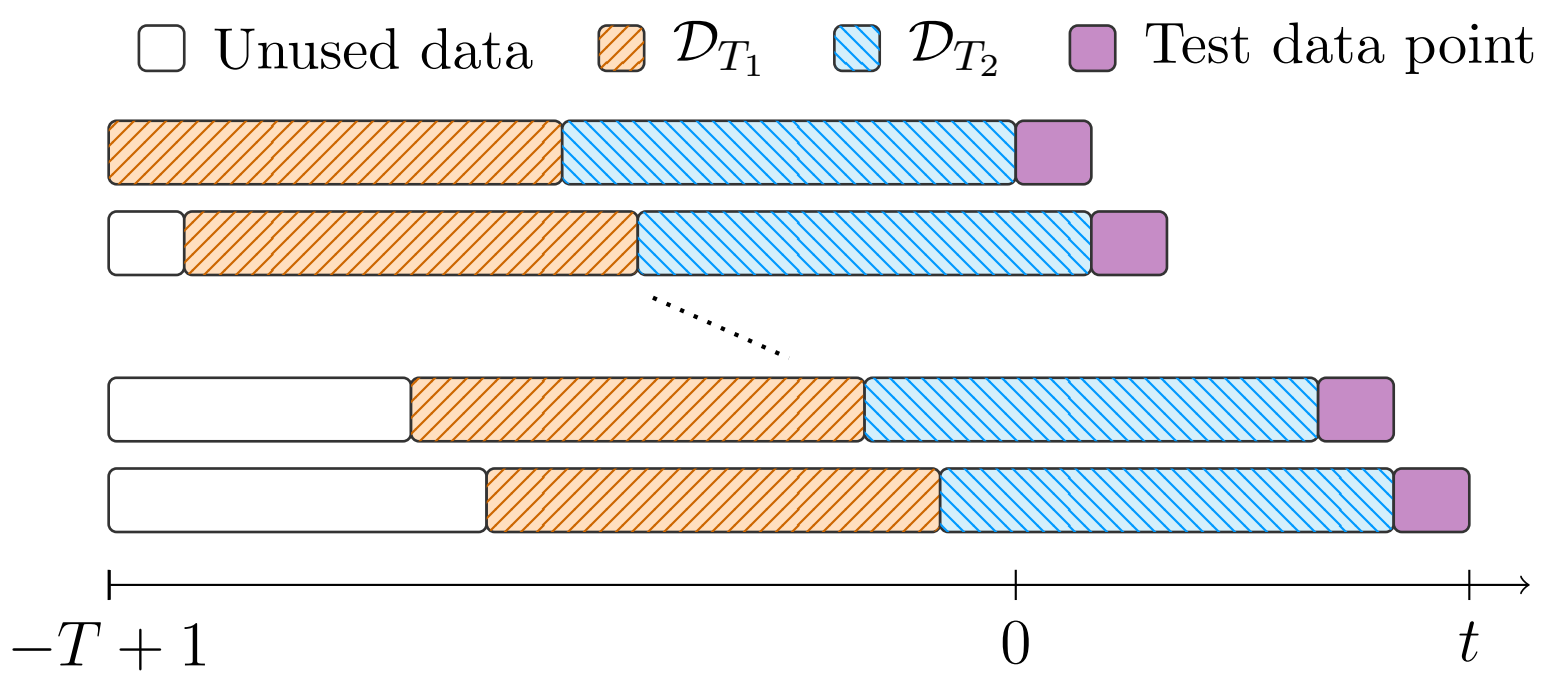

In the time series setting, we index observations by time $t$ instead of $i$, and the notation $n$, $n_1$, $n_2$ for full and split conformal prediction are replaced by $T$, $T_1$, $T_2$ for ACI. Following \citet{zaffran_2022}, we use Online Sequential Split Conformal Prediction (OSSCP) as the baseline design for ACI to adaptively update the miscoverage level. See Figure \ref{fig:osscp} for a scheme of OSSCP. 

\noindent Consider an initial training dataset $\mathcal{D}_T=\{(X_t,Y_t)\}_{t=-T+1}^{0}$ with $T$ observations, and a test data point $(X_{1}, Y_{1})$ with $Y_{1}$ not yet revealed. Similar to split conformal prediction, ACI splits the training dataset $\mathcal{D}_T$ into a proper training set $\mathcal{D}_{T_1}$ and a calibration set $\mathcal{D}_{T_2}$. A prediction model $\hat{f}_{T_1}$ is fitted on the proper training set $\mathcal{D}_{T_1}$ and used to compute nonconformity scores on the calibration set $\mathcal{D}_{T_2}$. A prediction set for $Y_{1}$ is then constructed as in \eqref{eq:split.conf.pred.set} with $T$, $T_1$ and $T_2$ in place of $n$, $n_1$ and $n_2$. Once $Y_{1}$ is revealed, the window for training dataset moves forward by one step, so the new training dataset is $\mathcal{D}_T=\{(X_t,Y_t)\}_{t=-T+2}^{1}$ and the new test data point is $(X_{2}, Y_{2})$ with $Y_{2}$ not yet revealed. With a slight abuse of notation, we keep writing $\mathcal{D}_T$ for the current training dataset, whose size stays fixed while its observations change over time. The same applies to $\mathcal{D}_{T_1}$, $\mathcal{D}_{T_2}$ and $\hat{f}_{T_1}$. The prediction set is thus updated sequentially. 

At any time $t \ge 1$, ACI updates the miscoverage level recursively with $\alpha_{1}=\alpha$:
\begin{align}\label{eq:aci}
\begin{cases}
\mathcal{C}^{\mathrm{ACI}}_{t-1,\alpha_t}(X_t)=\left [\hat f_{T_1}(X_t) \pm \widehat Q\left((1-\alpha_t)(1+\frac{1}{T_2})\right)\right ] \\
\alpha_{t+1}=\alpha_t+\gamma (\alpha-\mathbbm{1}\{Y_t \not \in \mathcal{C}^{\mathrm{ACI}}_{t-1,\alpha_t}(X_t)\}),
\end{cases}
\end{align}
where $\mathcal{C}_{t-1,\alpha_t}^{\mathrm{ACI}}$ uses the absolute residual as score, and takes the same form as \eqref{eq:pred.set.absolute.resid}, $\gamma > 0$ is a learning rate controlling the adaptation to distributional shifts, and $\alpha$ is the target level. The subscript $\alpha_t$ in $\mathcal{C}_{t-1,\alpha_t}$ indicates the miscoverage level used at time $t$, and the subscript $t-1$ denotes that it is constructed from the data available up to time $t-1$. If the prediction set $\mathcal{C}^{\mathrm{ACI}}_{t-1,\alpha_t}(X_t)$ covers the observed response $Y_t$ at time $t$, the miscoverage level increases at the next time point, potentially shrinking the next prediction set; otherwise, the next prediction set is potentially enlarged. As defined in \eqref{eq:aci}, $\alpha_t$ can fall outside $[0,1]$, in which case the prediction set is the whole real line $\mathbb{R}$ (if $\alpha_t\le0$) or empty (if $\alpha_t\ge1$). By adjusting the miscoverage level to the past coverage error, ACI adapts itself to unknown distributional shifts.

The price of adapting to unknown distributional shifts is that ACI does not guarantee the per-step marginal validity as in \eqref{eq:marginal_validity} or \eqref{eq:marginal_validity_split} for full or split conformal prediction. Instead, it controls the \textit{realized} average miscoverage rate over time in the long run. The inequality in Proposition \ref{prop:aci.conv} follows by summing both sides of the second equation in \eqref{eq:aci} up to any stopping time $\tau \ge1$, together with the fact that $\alpha_{\tau+1}>-\gamma$. Proposition 4.1 in \cite{gibbs_candes_2021} states a two-sided version of the inequality: $|(1/\tau)\sum_{t=1}^\tau Z_t -\alpha| \le (\max\{\alpha,1-\alpha\}+\gamma)/(\tau\gamma)$, which implies that the realized average miscoverage rate converges to the target level $\alpha$, on every sample path. In this paper, however, we only need the one-sided inequality in Proposition \ref{prop:aci.conv}, which has a much tighter upper bound for $\alpha<0.5$.

\begin{proposition}[Adapted from Proposition 4.1 in \citealp{gibbs_candes_2021}]\label{prop:aci.conv}
  For the updating procedure of ACI shown in \eqref{eq:aci}, with probability one we have that for all $\tau \ge 1$,
  \begin{equation*}
    \frac{1}{\tau}\sum_{t=1}^{\tau}Z_t\le\alpha+\frac{\alpha+\gamma}{\gamma\tau},
  \end{equation*}
  where $Z_t=\mathbbm{1}\{Y_t\notin\mathcal{C}^{\mathrm{ACI}}_{t-1,\alpha_t}(X_t)\}$.
\end{proposition}

Despite the fact that ACI adapts to unknown distributional shifts, its validity guarantee has two limitations. First, it only controls the realized average miscoverage rate, not the true average miscoverage rate, which is often more statistically relevant. The latter characterizes the statistical reliability of prediction sets in terms of their expected coverage performance, rather than the empirical coverage observed over a particular sequence of observations. Second, ACI only guarantees that the realized average approaches the target level $\alpha$ in the long run, and the excess term $(\alpha+\gamma)/(\gamma\tau)$ over $\alpha$ can be large for small and moderate $\tau$, potentially making the realized average exceed $\alpha$ over short horizons. In this paper, we address the limitations using anytime valid confidence sequences.

\section{Set-up}\label{sec:setup}


We follow the set-up of ACI in Section \ref{ssec:aci}. The observations $(X_t,Y_t)$, where $Y_t\in\mathcal{Y}=\mathbb{R}$ denotes the response and $X_t \in \mathcal{X}=\mathbb{R}^d$ denotes a $d$-dimensional feature vector, arrive sequentially over time. The first $T$ observations, indexed by $t=-T+1,\dots,0$, constitute the initial training dataset, and the online prediction begins at $t=1$. For each $t\ge1$, let $\mathcal{D}_{t-1}=\{(X_s,Y_s)\}_{s=-T+1}^{t-1}$ denote the collection of all observations available prior to time $t$, including the initial training dataset. Following the OSSCP design in Figure \ref{fig:osscp}, at each time $t\ge1$, the training dataset consists of the most recent $T$ observations in $\mathcal{D}_{t-1}$, which is further split into a proper training set of size $T_1$ and a calibration set of size $T_2$. 

Our methods in Section \ref{sec:theory} construct prediction sets of the same form as ACI in \eqref{eq:aci}, and differ only in the miscoverage level we plug in. ACI uses the level denoted as $\alpha_t$, and our methods in Section \ref{sec:theory} use a level denoted as $\eta_t$, which is computed from $\mathcal{D}_{t-1}$. In both cases, the plugged-in level is adjusted over time, while the target level $\alpha$ stays fixed and enters only through the updating rule. We therefore define the prediction set at time $t$ with the level $\eta_t$ as
\begin{equation*}
  \mathcal{C}_{t-1,\eta_t}(X_t)=\left[\hat f_{T_1}(X_t)\pm\widehat Q\left((1-\eta_t)(1+\frac{1}{T_2})\right)\right].
\end{equation*}
The prediction set shrinks as $\eta_t$ increases. The level $\eta_t$ can fall outside $[0,1]$, in which case the prediction set is the whole real line $\mathbb{R}$ (if $\eta_t\le0$) or empty (if $\eta_t\ge1$). For each prediction set, we track two quantities as in \eqref{eq:Z}: the miss indicator $Z_t$, which is observed once $Y_t$ is revealed, and the true miscoverage rate $\nu_t$, which is never observed. 
\begin{align}\label{eq:Z}
\begin{cases}
Z_t=\mathbbm{1}\{Y_t\notin\mathcal{C}_{t-1,\eta_t}(X_t)\} \\
\nu_t=\mathbb{P}\big(Y_t\notin\mathcal{C}_{t-1,\eta_t}(X_t)\mid\mathcal{F}_{t-1}\big)=\mathbb{E}[Z_t\mid\mathcal{F}_{t-1}],
\end{cases}
\end{align}
where $\mathcal{F}_{t-1}=\sigma(\mathcal{D}_{t-1},X_t)$ is the $\sigma$-field generated by all observations up to time $t-1$ and the current feature vector $X_t$, i.e., by all information available when the prediction set at time $t$ is constructed. We say that the procedure \textit{abstains} at time $t$ if it issues the whole real line $\mathbb{R}$ instead of a bounded prediction set. At an abstention, $Z_t=\nu_t=0$. The miss indicator $Z_t$ records whether the prediction set at time $t$ misses the particular value of the realized $Y_t$, while the true miscoverage rate $\nu_t$ is the probability that this prediction set misses over all values that $Y_t$ can take given all data available when the prediction set is constructed. Hence, $\nu_t$ describes the reliability of the prediction set itself, rather than the luck of a particular draw of $Y_t$. 

In \eqref{eq:Z}, we condition on $\mathcal{F}_{t-1}$, rather than directly $\mathcal{D}_{t-1}$. The $\sigma$-field represents all the information that can be extracted from the data. In stochastic processes, it is a convention to condition on the information available at each time. Since $\mathcal{F}_{t-1}=\sigma(\mathcal{D}_{t-1}, X_t)$ is exactly the information contained in $\mathcal{D}_{t-1}$ and $X_t$, the true miscoverage rate $\nu_t$ defined in \eqref{eq:Z} is the same quantity as defined in \eqref{eq:running} and \eqref{eq:rolling} in Section \ref{sec:intro}, except that it also conditions on the current feature vector $X_t$. Lemma \ref{lem:skip} and Theorems \ref{thm:run} and \ref{thm:roll} hold verbatim with $\mathcal{D}_{t-1}$ in place of $\mathcal{F}_{t-1}$, since whether a bounded prediction set is issued at time $t$ is determined by $\mathcal{D}_{t-1}$. In streaming time series, the information grows over time, i.e., $\mathcal{F}_{t-1}\subseteq\mathcal{F}_t$, and a filtration $(\mathcal{F}_t)_{t\ge0}$ records this sequential structure. 

\subsection{An Impossibility Result}\label{ssec:impossibility}

A natural goal is to uniformly control the true miscoverage rate of every single prediction set below the target level $\alpha$ at an arbitrary stopping time $\tau \ge 1$ with high probability, i.e., for a small $\delta >0$,
\begin{equation}\label{eq:pac_t}
  \mathbb{P}(\forall \tau \ge1: \nu_\tau \le \alpha)\ge 1-\delta. 
\end{equation}
This validity guarantee has been established for IID data \citep{avelin_2023,xu_karampatziakis_mineiro_2024,scharfstein_2026} and under distributional shift with a known importance weight \citep{hultberg_2026}. However, the importance weight is rarely known, and estimating it leads to errors that invalidate the finite-sample guarantee of conformal prediction. We argue that it is impossible to achieve this goal under potential distributional shifts without any further assumption on their nature, unless unbounded prediction sets are issued.

\begin{proposition}\label{prop:no.pac}
  Let $P$ be any underlying data-generating process, and $t^*\ge1$ be an arbitrary time point, at which a distributional shift may occur. A shifted process $P^*$ is a data-generating process under which $(\mathcal{D}_{t^*-1}, X_{t^*})$ has the same distribution as under $P$ but the distribution of the response $Y_{t^*}$ and of all subsequent observations may differ. Consider a procedure whose prediction set $\mathcal{C}_{t^*-1}(X_{t^*})$ is bounded with probability one under the underlying process $P$. Then, for any $\varepsilon\in(0,1)$, there exists a shifted process $P^*$ under which $\mathbb{P}(\nu_{t^*}\ge1-\varepsilon)\ge1-\varepsilon$.
\end{proposition}
Proposition~\ref{prop:no.pac} rules out the validity guarantee in \eqref{eq:pac_t} under unknown distributional shifts. For any $\alpha\in(0,1)$ and $\delta\in(0,1)$, choosing $\varepsilon<\min(1-\alpha,1-\delta)$ in Proposition~\ref{prop:no.pac} yields a shifted process $P^*$ under which $\mathbb{P}(\nu_{t^*} > \alpha) \ge \mathbb{P}(\nu_{t^*} \ge 1-\varepsilon) \ge 1-\varepsilon>\delta$. Hence,
\begin{equation*}
  \mathbb{P}(\forall \tau\ge1:\nu_\tau\le\alpha)\le\mathbb{P}(\nu_{t^*}\le\alpha)<1-\delta,
\end{equation*}
under the shifted process $P^*$. Consequently, no procedure that issues bounded prediction sets satisfies \eqref{eq:pac_t} under unknown distributional shifts. 

\subsection{Targets}\label{ssec:target}

As implied by Proposition \ref{prop:no.pac}, without any restrictive assumption on the nature of distributional shifts, the only way to control every single $\nu_t$ is to issue the whole real line, which is uninformative. We therefore target the averages of the true miscoverage rate $\nu_t$ instead of every single $\nu_t$. Specifically, we consider two average quantities:
\begin{alignat}{2}
  &\text{(Running)}\quad && \bar\nu_\tau=\frac{1}{\tau}\sum_{t=1}^{\tau}\nu_t,\label{eq:running_average}\\
  &\text{(Rolling)}\quad && \bar\nu^{(k)}_\tau=\frac{1}{k_\tau}\sum_{t=\tau-k_{\tau}+1}^{\tau}\nu_t,\label{eq:rolling_average}
\end{alignat}
where $\tau\ge1$ is an arbitrary stopping time, at which the average is evaluated, and $k_\tau\in\{1,\dots,\tau\}$ is the length of the rolling window, i.e., the number of most recent periods, over which the rolling average is measured. The running average is evaluated over the entire period elapsed since monitoring began, while the rolling average is measured over recent periods only, thereby reflecting the procedure's current reliability. 

A window rule is a deterministic sequence $(k_\tau)_{\tau\ge1}$ of window lengths, chosen before the monitoring starts, with $k_\tau-k_{\tau-1}\le1$, so that the window grows by at most one period per time step and the start of the window never moves backwards. The running average is the special case $k_\tau=\tau$, which is treated separately in Section \ref{ssec:running}. For the rolling average, we consider three window rules:
\begin{enumerate}[label={(\arabic*)}, align=left, leftmargin=*]
    \setlength\itemsep{-0.2em}
    \item proportional window $k_\tau=\lceil\theta\tau\rceil$ with $\theta\in(0,1)$, which covers a fixed fraction of the history;
    \item polynomial window $k_\tau=\lceil\tau^\beta\rceil$ with $\beta\in(1/2,1)$, which grows, but more slowly than the time;
    \item fixed window $k_\tau=\min\{\tau,k\}$, which covers the last $k$ periods.
\end{enumerate}

As implied by Proposition \ref{prop:no.pac}, the coverage at time $t^*$ at which a distributional shift may occur can fail, with $\nu_{t^*}$ close to one. However, such a single coverage failure raises the running average by at most $1/\tau$ and the rolling average by at most $1/k_\tau$, and can therefore be absorbed if the average was below $\alpha$ with enough room to spare. 

\subsection{Inferential Goals}

Our inferential goals are the validity guarantees on the true running and rolling average miscoverage rates. For a small $\delta >0$, 
\begin{alignat}{2}
  &\text{(Running)}\quad && \mathbb{P}\left(\forall\tau\ge1:\ \bar\nu_\tau\le\alpha\right)\ge1-\delta,\label{eq:running_goal}\\
  &\text{(Rolling)}\quad && \mathbb{P}\left(\forall\tau\ge1:\ \bar\nu^{(k)}_\tau\le\alpha\right)\ge1-\delta.\label{eq:rolling_goal}
\end{alignat}
Both guarantees hold at any, possibly data-dependent, stopping time. Moreover, they hold under unknown distributional shifts, without restrictions on their nature, e.g., timing, form, or magnitude.

\paragraph{Feasibility Check.}
To ensure that our methods in Sections \ref{ssec:running} and \ref{ssec:rolling} achieve the inferential goals, we need to perform a feasibility check such that a bounded prediction set is issued only if there is enough room below the target level $\alpha$ to absorb a miss at that time, as a distributional shift may occur at any time and cannot be anticipated.

Take the running average as an example. As implied by Proposition \ref{prop:no.pac}, if a distributional shift occurs at time $t^*$, the prediction set issued at that time misses with $\nu_{t^*}$ close to one, and the running average $\bar\nu_{t^*}$ becomes close to $(\sum_{t<t^*}\nu_t+1)/t^*$. Hence, a bounded prediction set at time $t^*$ is feasible only if 
\begin{equation}\label{eq:feasibility_running}
  \sum_{t<t^*}\nu_t+1\le\alpha t^*.
\end{equation}
Since the procedure decides whether to issue a bounded prediction set based on the data up to time $t^*-1$ only, it cannot know whether a shift occurs at time $t^*$. Consequently, a procedure that satisfies the validity guarantee for the running average under every data-generating process issues a bounded prediction set only if the condition in \eqref{eq:feasibility_running} holds, with probability at least $1-\delta$:
\begin{equation}\label{eq:feasible}
  \mathbb{P}\Big(\mathcal{C}_{t^*-1}(X_{t^*})\text{ is unbounded, or }\sum_{t<t^*}\nu_t+1\le\alpha t^*\Big)\ge1-\delta.
\end{equation}
In particular, the prediction sets before time $1/\alpha$ must be the whole real line $\mathcal{Y}=\mathbb{R}$ with probability at least $1-\delta$ (Appendix \ref{app:feasible.ineq}). Since $\nu_t$ is unknown, we replace $\sum_{t<t^*}\nu_t$ by its upper confidence bound and abstain whenever this upper bound plus one exceeds $\alpha t^*$. If the confidence sequence holds, which has probability at least $1-\delta$, a bounded set that passes this check is then feasible.

For the running average, no explicit feasibility check is needed: the buffer $B$ in \eqref{eq:sum.aci.certified} reserves enough room to absorb a coverage failure, so the check is passed whenever the miscoverage level $\eta_t$ is positive (Section \ref{ssec:running}). For the rolling average, the same condition must hold for every window that contains $t^*$, and an explicit feasibility check is required (Section \ref{ssec:rolling}).

\section{Theory and Methods}\label{sec:theory}

By definition of the miss indicator $Z_t$ and the true miscoverage rate $\nu_t$ in \eqref{eq:Z}, we have
  $\mathbb{E}[\nu_t - Z_t \mid \mathcal{F}_{t-1}]=0$,
which means that the differences $(\nu_t-Z_t)_{t\ge1}$ form a martingale difference sequence with respect to the filtration $(\mathcal{F}_t)_{t\ge0}$, regardless of potential distributional shifts. Also, the differences are bounded, since $\nu_t,Z_t\in[0,1]$. This allows us to apply existing anytime valid confidence sequences for sums of bounded martingale difference sequences \citep[cf.][]{howard_2020,howard_2021,waudby_2024} such that the sum of the true miscoverage rates $\sum_{t\le\tau}\nu_t$ is bounded from above by the sum of misses $\sum_{t\le\tau}Z_t$ plus a data-dependent margin, simultaneously for all periods up to an arbitrary stopping time with high probability; a two-sided confidence sequence additionally provides a lower bound. We apply the confidence sequence only at the steps at which the procedure issues a bounded prediction set, and skip the abstentions. At an abstention, $Z_t=\nu_t=0$, so skipping it changes neither the sum of misses nor the sum of true miscoverage rates, while including it would widen the margin without adding any information.

\begin{lemma}[Confidence sequences for true running average miscoverage rate]\label{lem:skip}
  Consider a procedure that, at each time $t$, decides to either issue a bounded prediction set or abstain, using only the data available up to time $t-1$. Apply an anytime valid confidence sequence to the true running average miscoverage rate in \eqref{eq:running_average}. 
  \begin{enumerate}[label={(\arabic*)}, align=left, leftmargin=*]
    \setlength\itemsep{-0.2em}
    \item With a one-sided confidence sequence, for a small $\delta>0$,
    \begin{equation}\label{eq:cs.run}
      \mathbb{P}\big(\forall\tau\ge1:\ \bar\nu_\tau\le\bar Z_\tau+\omega_\tau\big)\ge1-\delta,
    \end{equation}
    where $\bar Z_\tau=\sum_{t=1}^{\tau}Z_t/\tau$ is the realized running average miscoverage rate, and $\omega_\tau$ is the upper margin of the one-sided confidence sequence at error level $\delta$, computed from the misses at the steps with a bounded prediction set, with $\omega_0=0$.
    \item With a two-sided confidence sequence, for a small $\delta>0$,
    \begin{equation}\label{eq:cs.two}
      \mathbb{P}\big(\forall\tau\ge1:\ \bar\nu_\tau \in [\bar Z_\tau \pm \omega'_\tau] \big)\ge1-\delta,
    \end{equation}
    where $\omega'_\tau$ is the margin of a two-sided confidence sequence at error level $\delta$, also computed from the misses at the steps with a bounded prediction set, with $\omega'_0=0$. 
  \end{enumerate}
\end{lemma}

Lemma \ref{lem:skip} establishes one-sided and two-sided confidence sequences for the true running average miscoverage rate. Part (1) suffices to construct a procedure that satisfies the validity guarantee for the running average in \eqref{eq:running_goal}. For the rolling average in \eqref{eq:rolling_goal}, we need Part (2): since $k_\tau\bar\nu^{(k)}_\tau=\tau\bar\nu_\tau-s\,\bar\nu_s$, where $s=\tau-k_\tau$ is the start of the rolling window, an upper bound for the rolling average requires an upper bound for $\bar\nu_\tau$ and a lower bound for $\bar\nu_s$. Lemma \ref{lem:skip} thus provides the building block for constructing procedures that satisfy both validity guarantees, which we develop later in Sections \ref{ssec:running} and \ref{ssec:rolling}.

Any anytime valid confidence sequence can be applied in Lemma \ref{lem:skip} as long as it allows the conditional mean $\nu_t$ to change over time, as it does under potential distributional shifts. The choice does not affect the validity guarantee in \eqref{eq:cs.run} and \eqref{eq:cs.two}, but it does affect the efficiency of the prediction sets: since we need to bound $\bar{Z}_\tau+\omega_\tau$ from above by $\alpha$, a smaller $\omega_\tau$ leaves more room for $\bar{Z}_\tau$, which permits higher miscoverage levels and hence narrower prediction sets. Let $S_\tau$ be the sum of misses and $N_\tau$ the number of bounded prediction sets up to time $\tau$. Define $S_{t-1}/N_{t-1}=0$ when $N_{t-1}=0$. For the one-sided confidence sequence in \eqref{eq:cs.run}, we consider three margins:
\begin{enumerate}[label={(\arabic*)}, align=left, leftmargin=*]
  \setlength\itemsep{-0.2em}
  \item Stitched Empirical Bernstein Margin \citep[Theorem 4]{howard_2021}:
  \begin{equation}\label{eq:margin.stitched}
    \begin{aligned}
      \omega_\tau=\frac{1}{\tau}\Big\{&1.7\sqrt{(\widehat V_\tau\vee1)\big[(\log\log(2(\widehat V_\tau\vee1)))_++0.72\log(5.19/\delta)\big]}\\
      &+3.4(\log\log(2(\widehat V_\tau\vee1)))_++2.45\log(5.19/\delta)\Big\},
    \end{aligned}
  \end{equation}
  where $\widehat V_\tau$ is the sum of $(Z_t-S_{t-1}/N_{t-1})^2$ over the steps $t\le\tau$ with a bounded prediction set, $a\vee b=\max\{a,b\}$ and $(a)_+=\max\{a,0\}$. It adapts to empirical variability, but the last term does not, so $\tau\omega_\tau$ stays above $2.45\log(5.19/\delta)$ however small $\widehat V_\tau$ is.
  \item Hoeffding Margin:
  \begin{equation}\label{eq:margin.hoeffding}
    \omega_\tau=\frac{1.7}{\tau}\sqrt{(N_\tau/4\vee1)\big[(\log\log(2(N_\tau/4\vee1)))_++0.72\log(5.19/\delta)\big]}.
  \end{equation}
  It does not adapt to empirical variability, since it replaces the conditional variance $\nu_t(1-\nu_t)$ of each miss with its upper bound $1/4$, so it gains nothing from the fact that misses are rare.
  \item Normal Mixture Margin \citep[Proposition S2]{howard_2021}:
  \begin{equation}\label{eq:margin.normal}
    \omega_\tau=\frac{1}{\tau}\sqrt{2\big(N_\tau/4+\rho\big)\log\Big(\frac{1}{2\delta}\sqrt{\frac{N_\tau/4+\rho}{\rho}}+1\Big)},
  \end{equation}
  where $\rho>0$ is a tuning constant. Like the Hoeffding margin, it replaces the conditional variance $\nu_t(1-\nu_t)$ of each miss with its upper bound $1/4$, but it mixes over normal distributions instead of stitching, so that $\tau\omega_\tau$ grows as $O(\sqrt{N_\tau\log N_\tau})$ rather than $O(\sqrt{N_\tau\log\log N_\tau})$. 
\end{enumerate}
See Appendix \ref{app:margins} for derivations. For the two-sided confidence sequence in \eqref{eq:cs.two}, all three margins use $\delta/2$ in place of $\delta$. We compare the efficiency of the three margins in simulations.

\subsection{True Running Average Miscoverage Rate}\label{ssec:running}
As argued in Section \ref{sec:intro}, we leverage the adaptability of ACI to unknown distributional shifts in achieving our inferential goals. Combining Proposition \ref{prop:aci.conv} and the one-sided confidence sequence for the true running average miscoverage rate in \eqref{eq:cs.run} leads to:
\begin{equation}\label{eq:aci.cs}
  \mathbb{P}\left(\forall\tau\ge1:\ \bar\nu_\tau\le\alpha+\frac{\alpha+\gamma}{\gamma\tau}+\omega_\tau\right)\ge1-\delta .
\end{equation}
In comparison to the inferential goal in \eqref{eq:running_goal}, the upper bound in \eqref{eq:aci.cs} exceeds $\alpha$ by two additional terms: the excess $(\alpha+\gamma)/(\gamma\tau)$ of ACI, and the margin $\omega_\tau$ of the one-sided confidence sequence. To dispose of both terms, we propose a new updating rule of the miscoverage level. Summing both sides of the original ACI updating rule in \eqref{eq:aci} gives 
\begin{equation}\label{eq:aci.rule.sum}
  \alpha_{t+1}=\alpha+\gamma(\alpha t-S_t),
\end{equation}
where $S_t=\sum_{s=1}^t Z_s$, and $Z_s=\mathbbm{1}\{Y_s\notin\mathcal{C}^{\mathrm{ACI}}_{s-1,\alpha_s}(X_s)\}$. It follows from \eqref{eq:aci.rule.sum} that ACI steers the realized sum of misses $S_t$ towards the allowance $\alpha t$. The two additional terms in \eqref{eq:aci.cs} appear because $S_t$ may overshoot $\alpha t$ by up to $(\alpha+\gamma)/\gamma$ before ACI adapts, and because the scaled margin $t\omega_t$ comes on top of $S_t$. 

To remove both terms, our new updating rule uses a \textit{certified} sum of misses $S^c_t=S_t+t\omega_t$ and a buffer $B$. Let $S_0^c=0$, 
\begin{equation}\label{eq:running.new.rule}
  \eta_{t+1}=\alpha+\gamma(\alpha t-S^c_t-B)
\end{equation}
where $B \ge \alpha/\gamma-\alpha+c$ denotes a buffer reserved for the overshoot of ACI, and $c \ge 1$ bounds the maximum one-step increase of the certified sum of misses such that $S^c_t-S^c_{t-1}\le c$. The constant only depends on the choice of the anytime valid confidence sequence. By definition, $B>0$. The new rule in \eqref{eq:running.new.rule} can be written recursively as
\begin{equation}\label{eq:running.new.rule.rec}
  \eta_{t+1}=\eta_t+\gamma\big(\alpha-(S^c_t-S^c_{t-1})\big)=\eta_t+\gamma\big(\alpha-Z_t-(t\omega_t-(t-1)\omega_{t-1})\big).
\end{equation}
For the new rule, the miss indicator $Z_t$ in original ACI \eqref{eq:aci} is replaced by the increase of the certified sum, $S^c_t-S^c_{t-1}$, which adds the increase of the scaled margin, $t\omega_t-(t-1)\omega_{t-1}$, to $Z_t$. 

The new rule thus differs from original ACI in two ways. First, as in \eqref{eq:running.new.rule.rec}, each update of the miscoverage level is smaller than that of ACI by $\gamma (t\omega_t - (t-1)\omega_{t-1})$, which widens the prediction sets and leaves room for the scaled margin $t\omega_t$ in the allowance $\alpha t$. Second, as in \eqref{eq:running.new.rule}, the overshoot is absorbed by the buffer $B$, which enters only through the starting value $\eta_1=\alpha-\gamma B$ and lowers the miscoverage level by $\gamma B$ at every step.

In our new rule, the prediction set at time $t$ takes the same form as $\mathcal{C}^{\mathrm{ACI}}_{t-1,\alpha_t}(X_t)$ in \eqref{eq:aci} but with $\eta_t$ in place of $\alpha_t$. Hence, if $\eta_t\le0$, the prediction set covers all of $\mathcal{Y}=\mathbb{R}$, i.e., the procedure abstains, exactly as ACI does when $\alpha_t\le0$. At an abstention, $Z_t=0$ and the scaled margin $t\omega_t$ is unchanged, so that $S^c_t=S^c_{t-1}$, and \eqref{eq:running.new.rule.rec} gives $\eta_{t+1}=\eta_t+\gamma\alpha$. The miscoverage level thus rises at each abstention until $\eta_t>0$, when the procedure issues a bounded prediction set again. Algorithm \ref{alg:run} displays the anytime valid conformal prediction for the true running average miscoverage rate.

\begin{algorithm}[H]
\caption{Anytime Valid Conformal Prediction for Running Avg. (anytimeCP-Run)}
\label{alg:run}
\KwIn{Target level $\alpha\in(0,1)$; error level $\delta\in(0,1)$; learning rate $\gamma>0$; one-sided confidence sequence at error level $\delta$ whose certified sum increases by at most $c$ per step; buffer $B\ge\alpha/\gamma-\alpha+c$.}
\KwOut{Prediction sets $\{\mathcal{C}_{t-1}(X_t)\}_{t\ge1}$.}
\textbf{Initialize:} $N_0\leftarrow0$; $S_0\leftarrow0$; $S^c_0\leftarrow0$\;
\For{$t = 1, 2, 3, \dotsc$}{
  \tcp{before observing $Y_t$}
  $\eta_t\leftarrow\alpha+\gamma\{\alpha(t-1)-S^c_{t-1}-B\}$\;
  \eIf{$\eta_t>0$}{
    issue $\mathcal{C}_{t-1}(X_t)=\mathcal{C}_{t-1,\eta_t}(X_t)$; $N_t\leftarrow N_{t-1}+1$\;
  }{
    issue $\mathcal{C}_{t-1}(X_t)=\mathbb{R}$ (abstention); $N_t\leftarrow N_{t-1}$\;
  }
  \tcp{after observing $Y_t$}
  $Z_t\leftarrow\mathbbm{1}\{Y_t\notin\mathcal{C}_{t-1}(X_t)\}$; $S_t\leftarrow S_{t-1}+Z_t$\;
  $\omega_t\leftarrow$ margin in \eqref{eq:margin.stitched}--\eqref{eq:margin.normal} if $\eta_t>0$; otherwise $\omega_t\leftarrow(t-1)\omega_{t-1}/t$\;
  $S^c_t\leftarrow S_t+t\omega_t$\;
}
\end{algorithm}

In Algorithm \ref{alg:run}, the constant $c$ bounds $S^c_t-S^c_{t-1}$ for all $t$ and all sequences of misses. Since $S^c_t-S^c_{t-1}=Z_t+\{t\omega_t-(t-1)\omega_{t-1}\}$ and $Z_t\le1$, it suffices to bound the increase of the scaled margin, which is zero at an abstention and largest at the first bounded prediction set, where $t\omega_t$ jumps from zero to its starting value. Hence, the closed-form expressions of $c$ for the three margins are as follows.
\begin{enumerate}[label={(\arabic*)}, align=left, leftmargin=*]
  \setlength\itemsep{-0.2em}
  \item Stitched Empirical Bernstein Margin: $c=1+1.7\sqrt{0.72\log(5.19/\delta)}+2.45\log(5.19/\delta)$
  \item Hoeffding Margin: $c=1+1.7\sqrt{0.72\log(5.19/\delta)}$
  \item Normal Mixture Margin: $c=1+\sqrt{2(1/4+\rho)\log\{(2\delta)^{-1}\sqrt{(1/4+\rho)/\rho}+1\}}$
\end{enumerate}
See Appendix \ref{app:c} for derivations of $c$. As suggested above, $c$ depends only on the error level $\delta$ and, for the normal mixture margin, additionally on $\rho$, but not on $\alpha$, $\gamma$ or the data. Theorem \ref{thm:run} establishes the validity guarantee of Algorithm \ref{alg:run}.

\begin{theorem}\label{thm:run}
  Let $\{\mathcal{C}_{t-1}(X_t)\}_{t\ge1}$ be constructed by Algorithm \ref{alg:run} with any $\gamma>0$ and $B\ge\alpha/\gamma-\alpha+c$. Without any assumption on the data distribution:
  \begin{enumerate}[label={(\arabic*)}, align=left, leftmargin=*]
    \setlength\itemsep{-0.2em}
    \item $\mathbb{P}\left(\forall\tau\ge1:\ \bar\nu_\tau\le\alpha\right)\ge1-\delta$, even under unknown distributional shifts;
    \item on every sample path, $S^c_\tau\le\alpha\tau$, i.e., $\bar Z_\tau+\omega_\tau\le\alpha$, for all $\tau\ge1$; in particular, $\bar Z_\tau\le\alpha$;
    \item if $\tau\omega_\tau$ is non-decreasing in $\tau$, as for the margins in \eqref{eq:margin.stitched}--\eqref{eq:margin.normal}, then, on every sample path, $S^c_\tau>\alpha\tau-(B+(1-\alpha)/\gamma+\alpha)$ for all $\tau\ge1$.
  \end{enumerate}
\end{theorem}

Part (1) of Theorem \ref{thm:run} is our inferential goal for the true running average miscoverage rate in \eqref{eq:running_goal}. It holds at any stopping time and under arbitrary distributional shifts. It follows from Part (2) of this theorem and Part (1) of Lemma \ref{lem:skip}: with high probability, $\bar\nu_\tau\le\bar Z_\tau+\omega_\tau\le\alpha$ for all $\tau\ge1$.
Part (2) of Theorem \ref{thm:run} holds on every sample path: the certified sum never exceeds the allowance. It is the argument of Proposition \ref{prop:aci.conv} applied to the certified sum. If $\eta_t>0$, the new updating rule in \eqref{eq:running.new.rule} gives $S^c_{t-1}<\alpha(t-1)+\alpha/\gamma-B\le\alpha t-c$, so even a miss at time $t$ leaves $S^c_t\le S^c_{t-1}+c<\alpha t$; if $\eta_t\le0$, the procedure abstains and $S^c_t=S^c_{t-1}$. The buffer thus absorbs the overshoot of ACI. Since $c\ge1$, a bounded set is issued only if $S^c_{t-1}+1<\alpha t$, so the procedure passes the feasibility check in \eqref{eq:feasible} automatically. With the smallest buffer $B=\alpha/\gamma-\alpha+c$, it abstains at the start of monitoring until $t>c/\alpha$.

Part (3) of Theorem \ref{thm:run} shows that the procedure is not more conservative than necessary, which rules out trivial procedures, such as always abstaining. Its condition holds for all three margins \eqref{eq:margin.stitched}--\eqref{eq:margin.normal}, since $\tau\omega_\tau$ is a non-decreasing function of $\widehat V_\tau$ or $N_\tau$, neither of which decreases. The certified average $S_\tau^c/\tau=\bar Z_\tau+\omega_\tau$ stays within $(B+(1-\alpha)/\gamma+\alpha)/\tau$ of $\alpha$, because $\eta_t$ cannot rise far above one: an empty set misses for sure and then lowers $\eta_t$. Hence the realized running average is about $\alpha-\omega_\tau$, and no procedure that guarantees $\bar\nu_\tau\le\alpha$ through a confidence sequence can do substantially better.

\subsection{True Rolling Average Miscoverage Rate}\label{ssec:rolling}

To achieve the inferential goal in \eqref{eq:rolling_goal} for the true rolling average miscoverage rate, we follow the same steps as in Section \ref{ssec:running}, with the window of length $k_t$ in place of the whole history. Let $s_t=t-k_t$ denote the start of the window ending at time $t$. The window thus contains the time points $s_t+1,\dots,t$. Summing the original ACI updating rule in \eqref{eq:aci} over the window gives
\begin{equation}\label{eq:aci.rule.window}
  \alpha_{t+1}=\alpha_{s_t+1}+\gamma\big(\alpha k_t-(S_t-S_{s_t})\big).
\end{equation}
It follows from \eqref{eq:aci.rule.window} that ACI steers the realized sum of misses in the window, $S_t-S_{s_t}$, towards the allowance $\alpha k_t$. By Proposition \ref{prop:aci.window}, the realized sum of misses, $S_t-S_{s_t}$, overshoots the allowance by less than $(1+\gamma(1+\alpha))/\gamma$. This is larger than the $(\alpha+\gamma)/\gamma$ of Proposition \ref{prop:aci.conv}, because the miscoverage level at the start of the window, $\alpha_{s_t+1}$, can be as large as $1+\gamma\alpha$, whereas the running sum starts at $\alpha_1=\alpha$.  

\begin{proposition}[ACI over windows]\label{prop:aci.window}
  For the original ACI shown in \eqref{eq:aci} and any window rule $(k_t)$, with probability one we have that for all $\tau\ge1$,
  \begin{equation*}
    \bar{Z}_\tau^{(k)}=\frac{S_\tau-S_{s_\tau}}{k_\tau}\le\alpha+\frac{1+\gamma(1+\alpha)}{\gamma k_\tau},
  \end{equation*}
  where
  $\bar{Z}_\tau^{(k)}=\frac{1}{k_\tau}\sum_{t=s_\tau+1}^{\tau}Z_t$ and $Z_t=\mathbbm{1}\{Y_t\notin\mathcal{C}^{\mathrm{ACI}}_{t-1,\alpha_t}(X_t)\}$.
\end{proposition}

Proposition \ref{prop:aci.window} establishes the upper bound for the realized rolling average miscoverage rate, $\bar Z^{(k)}_\tau$, over the window, but the inferential goal in \eqref{eq:rolling_goal} concerns the true rolling average, $\bar\nu^{(k)}_\tau$. Combining the definition of $\bar\nu^{(k)}_\tau$ in \eqref{eq:rolling_average} and the two-sided confidence sequence in \eqref{eq:cs.two} leads to
\begin{equation}\label{eq:cs.roll}
  \mathbb{P}\left(\forall\tau\ge1:\ \bar\nu^{(k)}_\tau\le\bar Z^{(k)}_\tau+\frac{\tau\omega'_\tau+s_\tau\omega'_{s_\tau}}{k_\tau}\right)\ge1-\delta.
\end{equation}
See Appendix \ref{app:cs.roll} for the derivation of \eqref{eq:cs.roll}. Combining \eqref{eq:cs.roll} and Proposition \ref{prop:aci.window} gives
\begin{equation}\label{eq:aci.cs.roll}
  \mathbb{P}\left(\forall\tau\ge1:\ \bar\nu^{(k)}_\tau\le\alpha+\frac{1+\gamma(1+\alpha)}{\gamma k_\tau}+\frac{\tau\omega'_\tau+s_\tau\omega'_{s_\tau}}{k_\tau}\right)\ge1-\delta .
\end{equation}
Similar to Section \ref{ssec:running}, the upper bound of $\bar{\nu}_\tau^{(k)}$ in \eqref{eq:aci.cs.roll} exceeds $\alpha$ by two additional terms: the excess $(1+\gamma(1+\alpha))/(\gamma k_\tau)$ of ACI, and the term $(\tau\omega'_\tau+s_\tau\omega'_{s_\tau})/k_\tau$ which now carries the margin at both ends of the window.

To remove both terms, we replace the realized sum of misses in the window, $S_t-S_{s_t}$, in \eqref{eq:aci.rule.window} by a \textit{certified} sum of misses over the window, $C(s_t,t)$, and a feasibility check (Remark \ref{rmk:feasible.roll} and Appendix \ref{app:feasible}). Let
\begin{equation}\label{eq:certified.sum.roll}
  C(s_t,t)=\min\{N_t-N_{s_t},\ S_t-S_{s_t}+t\omega'_t+s_t\,\omega'_{s_t}\},
\end{equation}
where the first term is the number of bounded sets in the window, which bounds $\sum_{i=s_t+1}^{t}\nu_i$ trivially since $\nu_i\le1$ at steps with a bounded set and $\nu_i=0$ at abstentions, and the second term bounds it with high probability, by \eqref{eq:cs.two}. With $C(s_0,0)=0$ and $k_0=0$, our new updating rule for a rolling window is
\begin{equation}\label{eq:eta.roll}
  \eta_{t+1}=\alpha+\gamma\big(\alpha k_t-C(s_t,t)-B\big).
\end{equation}
For the running average, the buffer $B$ in \eqref{eq:running.new.rule} is a fixed reserve that replaces a feasibility check, since the certified sum changes only at steps with a bounded set, and by at most $c$. For a rolling window, no fixed reserve suffices, because the certified sum can also grow at abstentions as the window start moves. Validity is therefore guaranteed by a specific feasibility check, and the buffer $B\ge0$ in \eqref{eq:eta.roll} only affects efficiency: with a small $B$, the certified sum stays close to the allowance and the check often fails, whereas a large $B$ lowers $\eta_t$, which widens the sets or leads to abstentions. The new rule in \eqref{eq:eta.roll} can be written recursively as
\begin{equation}\label{eq:eta.roll.rec}
  \eta_{t+1}=\eta_t+\gamma\big(\alpha(k_t-k_{t-1})-(C(s_t,t)-C(s_{t-1},t-1))\big).
\end{equation}
Unlike the running window which keeps old misses forever, a rolling window forgets them: when the rolling window start moves past a miss, the certified sum falls, so $\eta_{t+1}$ rises and the prediction sets become smaller. As in Section \ref{ssec:running}, the procedure abstains if $\eta_t\le0$. 

\begin{remark}[Feasibility check]\label{rmk:feasible.roll}
  For the running average, the certified sum changes only at steps with a bounded set. For a rolling window, however, the certified sum can also increase at abstentions: when the window start moves from $s$ to $s'>s$, the margin $s'\omega'_{s'}$ at the new start may exceed $s\,\omega'_s$. Algorithm \ref{alg:roll} therefore issues a bounded set at time $t$ only if, whatever the value of $Z_t$, every later window that contains $t$ stays within its allowance even if all later steps abstain. Such a window starts at some $s\in\{s_t,\dots,t-1\}$, and the check is $C^{(z)}(s,t)\le\alpha\,m_t(s)$ for $z\in\{0,1\}$, where $C^{(z)}(s,t)$ is the certified sum of the window after $s$, computed as if a bounded set were issued at $t$ with $Z_t=z$, and $m_t(s)=\max(t,\tau_s)-s$, with $\tau_s=\min\{u:s_u=s\}$, is its smallest length. 
\end{remark}

\begin{algorithm}[H]
\caption{Anytime Valid Conformal Prediction for Rolling Avg. (anytimeCP-Roll)}
\label{alg:roll}
\KwIn{Target level $\alpha\in(0,1)$; error level $\delta\in(0,1)$; learning rate $\gamma>0$; buffer $B\ge0$; window rule $(k_t)$; two-sided confidence sequence at error level $\delta$.}
\KwOut{Prediction sets $\{\mathcal{C}_{t-1}(X_t)\}_{t\ge1}$.}
\textbf{Initialize:} $N_0\leftarrow0$; $S_0\leftarrow0$; $k_0\leftarrow0$; $C(s_0,0)\leftarrow0$\;
\For{$t = 1, 2, 3, \dotsc$}{
  \tcp{before observing $Y_t$}
  $\eta_t\leftarrow\alpha+\gamma\big(\alpha k_{t-1}-C(s_{t-1},t-1)-B\big)$\;
  \tcp{feasibility check}
  \For{$z\in\{0,1\}$ and $s\in\{s_t,\dots,t-1\}$ with $\tau_s<\infty$}{
    $\omega'^{(z)}_t\leftarrow$ margin at error level $\delta/2$ as if a bounded set were issued at $t$ with $Z_t=z$\;
    $C^{(z)}(s,t)\leftarrow\min\{N_{t-1}+1-N_s,\ S_{t-1}+z-S_s+t\,\omega'^{(z)}_t+s\,\omega'_s\}$\;
  }
  \eIf{$\eta_t>0$ and $C^{(z)}(s,t)\le\alpha\,m_t(s)$ for all $z$ and $s$ above}{
    \tcp{check passed}
    issue $\mathcal{C}_{t-1}(X_t)=\mathcal{C}_{t-1,\eta_t}(X_t)$; $N_t\leftarrow N_{t-1}+1$\;
  }{
    \tcp{$\eta_t\le0$ or check failed}
    issue $\mathcal{C}_{t-1}(X_t)=\mathbb{R}$ (abstention); $N_t\leftarrow N_{t-1}$\;
  }
  \tcp{after observing $Y_t$}
  $Z_t\leftarrow\mathbbm{1}\{Y_t\notin\mathcal{C}_{t-1}(X_t)\}$; $S_t\leftarrow S_{t-1}+Z_t$\;
  $\omega'_t\leftarrow$ margin \eqref{eq:margin.stitched}--\eqref{eq:margin.normal} at level $\delta/2$ if $\mathcal{C}_{t-1}(X_t)\neq\mathbb{R}$; otherwise $\omega'_t\leftarrow(t-1)\omega'_{t-1}/t$\;
  $C(s_t,t)\leftarrow\min\{N_t-N_{s_t},\ S_t-S_{s_t}+t\,\omega'_t+s_t\,\omega'_{s_t}\}$\;
}
\end{algorithm}

\begin{theorem}\label{thm:roll}
  Let $\{\mathcal{C}_{t-1}(X_t)\}_{t\ge1}$ be constructed by Algorithm \ref{alg:roll} for a window rule $(k_t)$, any $\gamma>0$ and any $B\ge0$. Without any assumption on the data distribution:
  \begin{enumerate}[label={(\arabic*)}, align=left, leftmargin=*]
    \setlength\itemsep{-0.2em}
    \item $\mathbb{P}\left(\forall\tau\ge1:\ \bar\nu^{(k)}_\tau\le\alpha\right)\ge1-\delta$, even under unknown distributional shifts;
    \item on every sample path, $C(s_\tau,\tau)\le\alpha k_\tau$ for all $\tau\ge1$; in particular, $\bar Z^{(k)}_\tau\le\alpha$.
  \end{enumerate}
\end{theorem}

Theorem \ref{thm:roll} establishes the validity guarantee of Algorithm \ref{alg:roll}. Part (1) of Theorem \ref{thm:roll} is our inferential goal in \eqref{eq:rolling_goal}. It holds for every window rule, at any stopping time and under arbitrary distributional shifts. It follows from Part (2) of this theorem and Part (2) of Lemma \ref{lem:skip}: with high probability, $\sum_{t=s_\tau+1}^{\tau}\nu_t\le C(s_\tau,\tau)\le\alpha k_\tau$ for all $\tau\ge1$. Part (2) of Theorem \ref{thm:roll} holds on every sample path: the certified sum in the window never exceeds the allowance. It follows from the feasibility check, since a bounded set is issued only if every window that contains it stays within its allowance, and since abstentions do not increase these certified sums. 

Theorem \ref{thm:roll} has no counterpart of Part (3) of Theorem \ref{thm:run}, because how often the procedure can issue bounded sets depends on the window rule. Validity holds for every window rule, but the certified sum exceeds the realized sum by up to the scaled margins at both ends of the window, $\tau\omega'_\tau+s_\tau\omega'_{s_\tau}$, which grow with $\tau$. If they exceed the allowance $\alpha k_\tau$, the procedure keeps $C(s_\tau,\tau)\le\alpha k_\tau$ only through the first term in \eqref{eq:certified.sum.roll}, i.e., $N_\tau-N_{s_\tau}\le\alpha k_\tau$: it then issues at most $\alpha k_\tau$ bounded sets per window and abstains at least a fraction $1-\alpha$ of the time. Proposition \ref{prop:regimes} shows for which window rules these scaled margins stay small relative to the window length.

\begin{proposition}[Efficiency loss of the rolling guarantee]\label{prop:regimes}
  Let the margin be one of \eqref{eq:margin.stitched}--\eqref{eq:margin.normal} at error level $\delta/2$. On every sample path, $C(s_\tau,\tau)-(S_\tau-S_{s_\tau})\le h(\tau)$, where $h(\tau)=O(\sqrt{\tau\log\tau})$ is a deterministic function given in the proof. Hence $\{C(s_\tau,\tau)-(S_\tau-S_{s_\tau})\}/k_\tau\le h(\tau)/k_\tau$, and:
  \begin{enumerate}[label={(\arabic*)}, align=left, leftmargin=*]
    \setlength\itemsep{-0.2em}
    \item proportional windows $k_\tau=\lceil\theta\tau\rceil$, $h(\tau)/k_\tau=O\big(\sqrt{\log\tau/\tau}\big)$;
    \item polynomial windows $k_\tau=\lceil\tau^{\beta}\rceil$ with $\beta\in(1/2,1)$, $h(\tau)/k_\tau=O\big(\tau^{1/2-\beta}\sqrt{\log\tau}\big)$;
    \item windows with $k_\tau=O(\sqrt{\tau})$, including fixed windows, the bound $h(\tau)/k_\tau$ does not vanish.
  \end{enumerate}
\end{proposition}

By Proposition \ref{prop:regimes}, the efficiency loss of the rolling guarantee vanishes asymptotically for proportional and polynomial windows. For windows of length $O(\sqrt{\tau})$, including fixed windows, it does not, since by the law of the iterated logarithm, the scaled margin $\tau\omega'_\tau$ of any time-uniform confidence sequence is at least of order $\sqrt{\tau}$. We therefore recommend windows that grow faster than $\sqrt{\tau}$. A fixed window gives an informative guarantee only over a finite monitoring period.

\section{Simulations}\label{sec:simulations}

We generate the data based on a simple linear autoregressive time-series model with exogenous regressors
\begin{equation}\label{eq:sim.dgp}
  y_t=0.5 y_{t-1}+\beta_{1,t}x_{1,t}-2x_{2,t}+1.5x_{3,t}+\varepsilon_t,\qquad \varepsilon_t\sim N(\mu_t,\sigma_t^2),
\end{equation}
with the autocorrelated regressor $x_{1,t}=0.5x_{1,t-1}+u_t$, $u_t\sim N(0,1)$, and $x_{2,t}\sim N(2,4)$ and $x_{3,t}\sim U(0,2)$ independent over time. Since conformal prediction is known to be distribution-free and model-agnostic, a simple linear model in \eqref{eq:sim.dgp} with little computational cost suffices to show the properties of our methods. More complicated data-generating models and more flexible ML prediction methods can be used instead without affecting the main conclusions.

Following the OSSCP scheme in Figure \ref{fig:osscp}, at each prediction step, a linear regression of $y_t$ on $y_{t-1}$, $x_{1,t}$, $x_{2,t}$ and $x_{3,t}$ is fitted on a proper training set of $1000$ observations, and absolute residuals are computed as scores on the subsequent rolling calibration set of $1000$ observations. There are $10^4$ test periods. To examine our algorithms, we study two scenarios:
\begin{equation*}
  \text{(Stable)}\ \ (\beta_{1,t},\mu_t,\sigma_t)=(3,0,1)\ \ \text{for all }t;
  \qquad
  \text{(Shift)}\ \ (\beta_{1,t},\mu_t,\sigma_t)=
  \begin{cases}(3,0,1), & t\le4000,\\ (3,2,1), & 4000<t\le7000,\\ (-1,0,3), & t>7000.\end{cases}
\end{equation*}
In the Stable scenario, the three parameters $(\beta_{1,t},\mu_t,\sigma_t)$ remain constant over all time periods. In the Shift scenario, however, at least one of the three parameters changes at $t=4000$ and $t=7000$. In addition, we use $\alpha=\delta=0.05$ for our algorithms and run $500$ Monte Carlo replications. The results for the Stable scenario are reported in Appendix \ref{app:noshift}.

Since the underlying data-generating model in \eqref{eq:sim.dgp} is known, we can precisely compute the true miscoverage rate for every prediction set $[l_t,u_t]$ at time $t$. Since $y_t\mid\mathcal{F}_{t-1}\sim N(m_t+\mu_t,\sigma_t^2)$ with $m_t=0.5y_{t-1}+\beta_{1,t}x_{1,t}-2x_{2,t}+1.5x_{3,t}$,
\begin{equation*}
  \nu_t=\mathbb{P}(y_t<l_t\mid\mathcal{F}_{t-1})+\mathbb{P}(y_t>u_t\mid\mathcal{F}_{t-1})
  =\Phi\Big(\frac{l_t-m_t-\mu_t}{\sigma_t}\Big)+1-\Phi\Big(\frac{u_t-m_t-\mu_t}{\sigma_t}\Big),
\end{equation*}
with $\nu_t=0$ for an abstention ($l_t=-\infty$, $u_t=\infty$). To evaluate the validity guarantee of our algorithms, we depict the true running average miscoverage rate $\bar\nu_\tau$, or the true rolling average miscoverage rate $\bar\nu_\tau^{(k)}$ for every replication against the target level $\alpha$. Also, we report the median and $95\%$ quantile across all replications of the maximal true average miscoverage rate, $\bar\nu_\tau$ or $\bar\nu_\tau^{(k)}$, over all test periods. To evaluate the efficiency of our algorithms, we report the share of abstentions and the median width of the bounded prediction sets over each of the three phases: (I) $t \le 4000$, (II) $4000<t\le 7000$, and (III) $t>7000$. 

Our proposed procedures are: Algorithm \ref{alg:run} with stitched margin \eqref{eq:margin.stitched}, Hoeffding margin \eqref{eq:margin.hoeffding} and normal mixture margin \eqref{eq:margin.normal} with $\rho=20$, denoted as anytimeCP-Run (Stitched), (Hoeffding) and (Normal-Mixture), each with $B=\alpha/\gamma-\alpha+c$. Algorithm \ref{alg:roll} with stitched margin for proportional windows $k_\tau=\lceil\tau/2\rceil$, polynomial windows $k_\tau=\lceil\tau^{0.9}\rceil$ and fixed windows $k_\tau=\min\{\tau,500\}$, denoted as anytimeCP-Roll (Prop), (Poly) and (Fixed), each with $B=10$, chosen from a small pilot study. We compare our proposed procedures with three benchmarks: split conformal prediction at the fixed miscoverage level $\alpha$ (Split CP), parametric prediction interval $[\hat{y}_t\pm z_{1-\alpha/2}\hat{s}_t]$, where $z_{1-\alpha/2}$ is the $(1-\alpha/2)$-quantile of the standard normal distribution, and $\hat{s}_t$ is the root mean squared error of calibration set at time $t$ (Parametric PI), and ACI. All methods use the same point predictions and calibration scores. ACI and our algorithms use the same learning rate $\gamma=0.01$. 

\subsection{True Running Average Miscoverage Rate}\label{ssec:sim.run}

\begin{figure}[!htbp]
\centering
\maybeinclude[width=0.85\textwidth]{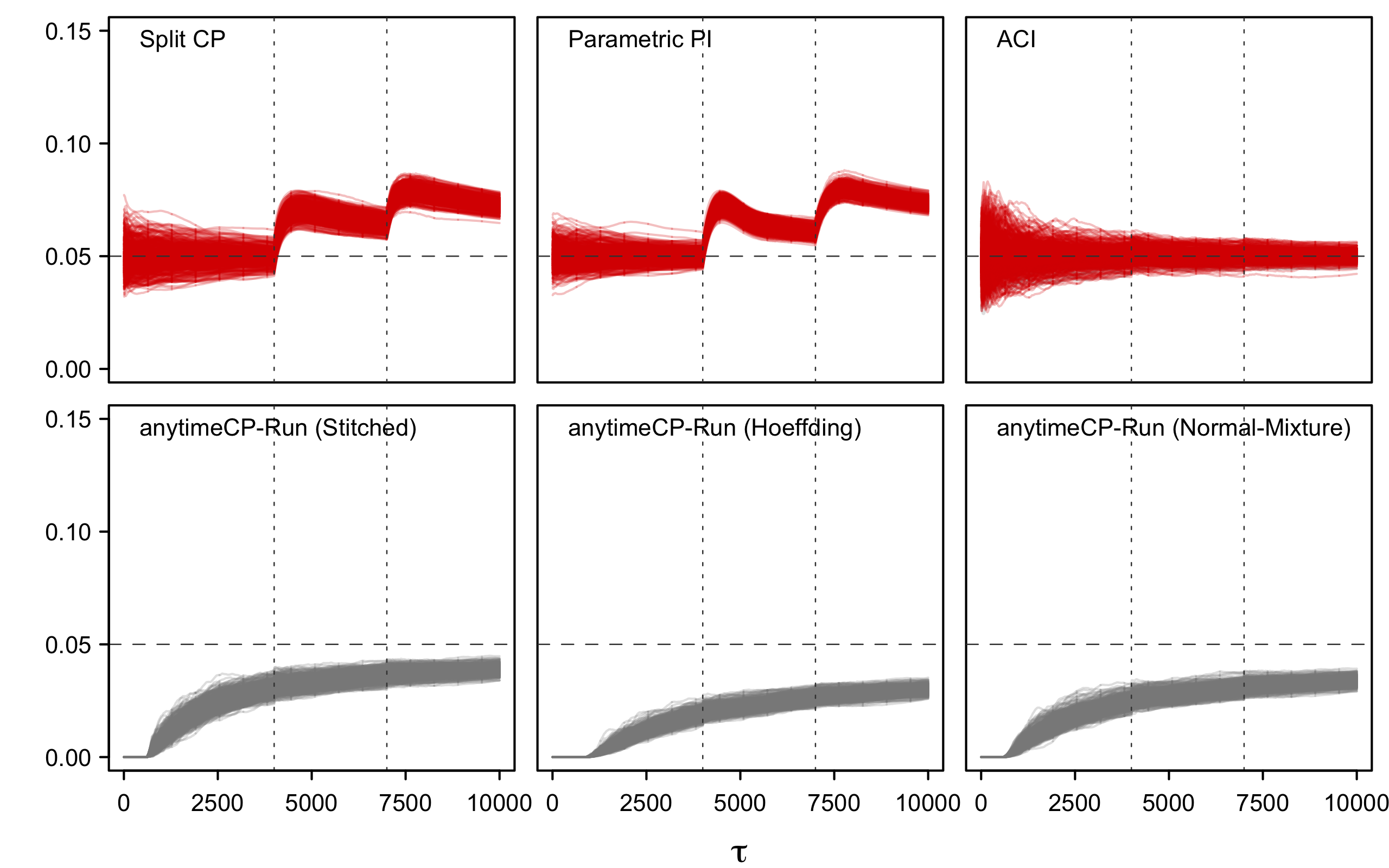}
\caption{True running average miscoverage rate $\bar\nu_\tau$ of each replication with distributional shifts at $\tau=4000$ and $\tau=7000$ (vertical lines). Red: paths that exceed $\alpha$ at some $\tau$. Grey: paths that do not.}
\label{fig:sim.run.paths}
\end{figure}

Figure \ref{fig:sim.run.paths} displays the true running average miscoverage rate of every replication. Split CP and Parametric PI exceed the target level $\alpha$ in every replication: before the first shift, their true running average fluctuates around $\alpha$, and exceeds it at some time in almost every replication; after the shifts, it rises to about $0.08$. Parametric PI performs similarly to Split CP, because the linear model with normal errors is correctly specified. ACI keeps the realized running average close to $\alpha$ (Proposition \ref{prop:aci.conv}), but its true running average also exceeds $\alpha$ in almost all replications. In comparison, the true running average miscoverage rate of our procedure in Algorithm \ref{alg:run} stays below $\alpha$ in every replication and for every type of margin, in line with Theorem \ref{thm:run}. It approaches $\alpha$ from below over time, since the margin of the anytime valid confidence sequence shrinks. As seen in Table \ref{tab:sim.run}, for our procedure with stitched margin, the median across all replications of the maximal true running average over all test periods is 0.039, below 0.05, whereas, for the three benchmarks, the medians are all well above 0.05.

Table \ref{tab:sim.run} also reports the efficiency loss of our anytime validity guarantee by the share of abstentions and the median width of prediction sets. Our procedure abstains mostly at the start. In Algorithm \ref{alg:run}, the level $\eta_t>0$ only once the allowance $\alpha(t-1)$  exceeds the certified sum plus the buffer, which takes a few hundred test periods. The median width of bounded prediction sets for our procedure is also larger than that for the three benchmark methods, because our procedure takes into account both the anytime validity guarantee and the unknown distributional shifts. The relative difference in median width, however, shrinks over time. Among the three margins, the stitched margin gives the narrowest prediction sets and the Hoeffding margin the widest, since only the stitched margin adapts to the small empirical variance of the rare misses, resulting from a small target level $\alpha=0.05$.

\begin{table}[!htbp]
\centering\spacingset{1}\footnotesize
\setlength{\tabcolsep}{4pt}\renewcommand{\arraystretch}{1.0}
\begin{tabular}{lcccccccc}
\toprule
& \multicolumn{2}{c}{$\max_{\tau\ge 1}\bar\nu_\tau$} & \multicolumn{3}{c}{Abstention} & \multicolumn{3}{c}{Width} \\
\cmidrule(lr){2-3}\cmidrule(lr){4-6}\cmidrule(lr){7-9}
& median & $95\%$ & I & II & III & I & II & III \\
\midrule
Split CP & 0.079 & 0.084 & 0 & 0 & 0 & 3.93 & 5.58 & 13.16 \\
Parametric PI & 0.080 & 0.083 & 0 & 0 & 0 & 3.93 & 5.57 & 12.63 \\
ACI & 0.059 & 0.073 & 0 & 2\% & 2\% & 3.94 & 6.04 & 14.75 \\
anytimeCP-Run (Stitched) & 0.039 & 0.042 & 16\% & 3\% & 3\% & 4.21 & 6.20 & 15.07 \\
anytimeCP-Run (Hoeffding) & 0.030 & 0.033 & 23\% & 3\% & 3\% & 4.54 & 6.42 & 15.59 \\
anytimeCP-Run (Normal-Mixture) & 0.034 & 0.037 & 15\% & 3\% & 3\% & 4.40 & 6.33 & 15.42 \\
\bottomrule
\end{tabular}
\caption{Summary statistics of true running average miscoverage rate for all replications with distributional shifts at $\tau=4000$ and $\tau=7000$.}
\label{tab:sim.run}
\end{table}

\subsection{True Rolling Average Miscoverage Rate}\label{ssec:sim.roll}

\begin{figure}[!htbp]
\centering
\maybeinclude[width=0.85\textwidth]{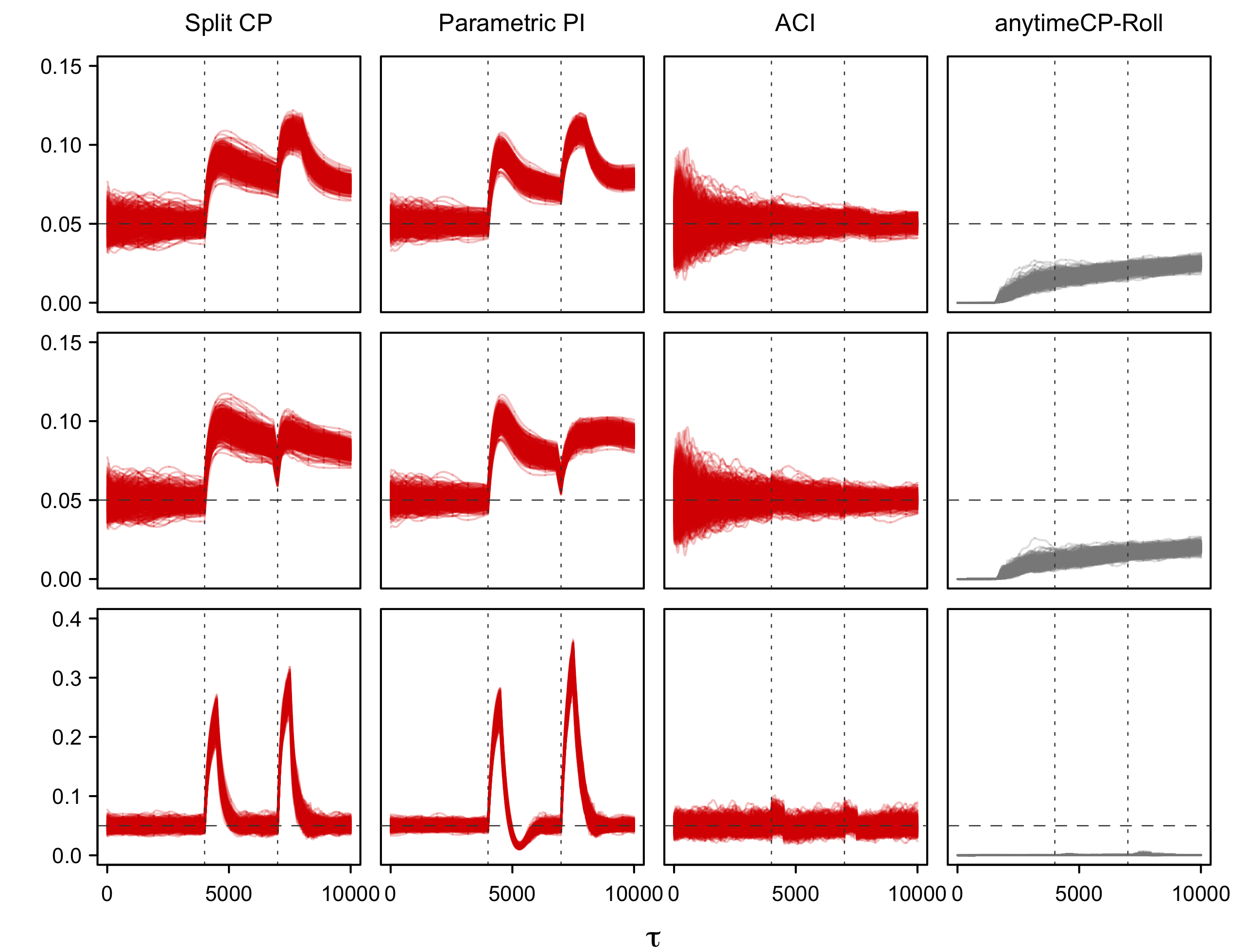}
\caption{True rolling average miscoverage rate $\bar\nu^{(k)}_\tau$ of each replication with distributional shifts at $\tau=4000$ and $\tau=7000$ (vertical lines). Rows: proportional windows $k_\tau=\lceil\tau/2\rceil$, polynomial windows $k_\tau=\lceil\tau^{0.9}\rceil$, fixed windows $k_\tau=\min\{\tau,500\}$. Red: paths that exceed $\alpha$ at some $\tau$; grey: paths that do not. }
\label{fig:sim.roll.paths}
\end{figure}

Figure \ref{fig:sim.roll.paths} displays the true rolling average miscoverage rate of every replication. Unlike the running average, a rolling window forgets the earlier test periods, so the sudden rise in the true miscoverage rate after a shift is averaged over fewer periods and is less diluted. Hence, the true rolling averages of the three benchmark methods rise further above $\alpha$, and much more so for fixed windows. The true rolling average miscoverage rate of our procedure in Algorithm \ref{alg:roll} stays below $\alpha$ in every replication and for every window rule, in line with Theorem \ref{thm:roll}. For proportional and polynomial windows, it approaches $\alpha$ from below over time, because the margin of the anytime valid confidence sequence shrinks. As shown in Table \ref{tab:sim.roll}, for our procedure with proportional windows, the median across all replications of the maximal true average over all test periods is 0.025, well below 0.05, whereas, for the three benchmarks, the medians are all well above 0.05.

\begin{table}[!htbp]
\centering\spacingset{1}\footnotesize
\setlength{\tabcolsep}{4pt}\renewcommand{\arraystretch}{1.0}
\begin{tabular}{llcccccccc}
\toprule
& & \multicolumn{2}{c}{$\max_{\tau\ge1}\bar\nu^{(k)}_\tau$} & \multicolumn{3}{c}{Abstention} & \multicolumn{3}{c}{Width} \\
\cmidrule(lr){3-4}\cmidrule(lr){5-7}\cmidrule(lr){8-10}
Window & Method & median & $95\%$ & I & II & III & I & II & III \\
\midrule
proportional & Split CP & 0.108 & 0.115 & 0 & 0 & 0 & 3.93 & 5.58 & 13.16 \\
 & Parametric PI & 0.110 & 0.116 & 0 & 0 & 0 & 3.93 & 5.57 & 12.63 \\
 & ACI & 0.067 & 0.083 & 0 & 2\% & 2\% & 3.94 & 6.04 & 14.75 \\
 & anytimeCP-Roll (Prop) & 0.025 & 0.029 & 46\% & 9\% & 7\% & 4.91 & 6.88 & 16.64 \\
\midrule
polynomial & Split CP & 0.098 & 0.108 & 0 & 0 & 0 & 3.93 & 5.58 & 13.16 \\
 & Parametric PI & 0.101 & 0.110 & 0 & 0 & 0 & 3.93 & 5.57 & 12.63 \\
 & ACI & 0.066 & 0.081 & 0 & 2\% & 2\% & 3.94 & 6.04 & 14.75 \\
 & anytimeCP-Roll (Poly) & 0.021 & 0.024 & 52\% & 13\% & 10\% & 5.07 & 7.11 & 17.25 \\
\midrule
fixed & Split CP & 0.279 & 0.302 & 0 & 0 & 0 & 3.93 & 5.58 & 13.16 \\
 & Parametric PI & 0.327 & 0.351 & 0 & 0 & 0 & 3.93 & 5.57 & 12.63 \\
 & ACI & 0.075 & 0.089 & 0 & 2\% & 2\% & 3.94 & 6.04 & 14.75 \\
 & anytimeCP-Roll (Fixed) & 0.001 & 0.003 & 96\% & 96\% & 96\% & 5.19 & 7.31 & 18.70 \\
\bottomrule
\end{tabular}
\caption{Summary statistics of true rolling average miscoverage rate for all replications with distributional shifts at $\tau=4000$ and $\tau=7000$.}
\label{tab:sim.roll}
\end{table}

Table \ref{tab:sim.roll} reports the efficiency loss of our anytime validity guarantee for the true rolling average, which depends on the window rule. For proportional and polynomial windows, our procedure in Algorithm \ref{alg:roll} abstains mostly at the start, until the window is long enough for the allowance $\alpha k_\tau$ to exceed the scaled margins at both ends of the window. The prediction sets for the rolling average are wider than for the running average, because the certified sum in the window carries scaled margins at both ends, while its allowance $\alpha k_\tau$ is smaller. For fixed windows, the scaled margins exceed the allowance $\alpha k=25$, so our procedure issues at most about $\alpha k$ bounded sets per window and abstains at least a fraction $1-\alpha$ of the time: the validity guarantee still holds, but only through abstentions. These results are in line with Proposition \ref{prop:regimes}: the efficiency loss shrinks over time for proportional and polynomial windows, but not for fixed windows.

\section{Discussion}\label{sec:discussion}

This paper studies the construction of anytime valid conformal prediction sets on streaming time series under unknown distributional shifts. Proposition \ref{prop:no.pac} shows that the true miscoverage rate of every single prediction set cannot be uniformly controlled under distributional shifts unless uninformative prediction sets are issued or the nature of such distributional shifts is assumed. We therefore target the true running and rolling average miscoverage rates. Our procedures combine ACI, which adapts to unknown distributional shifts, with anytime valid confidence sequences: ACI is applied to a certified sum of misses, i.e., the realized sum of misses plus the scaled margin of the confidence sequence, which bounds the true sum of misses with high probability. Algorithm \ref{alg:run} controls the true running average miscoverage rate at any stopping time (Theorem \ref{thm:run}). Algorithm \ref{alg:roll}, which adds a feasibility check, controls the true rolling average miscoverage rate for any window rule (Theorem \ref{thm:roll}). The rolling guarantee holds at every time for every window rule; its loss of efficiency, through abstentions and wider prediction sets, shrinks over time for proportional and polynomial windows, but not for fixed windows (Proposition \ref{prop:regimes}). In practice, we recommend the running average whenever a guarantee over the whole history is acceptable to the user, since it abstains less often and gives narrower prediction sets. If the guarantee should instead reflect the most recent periods, we recommend the rolling average with proportional or polynomial windows; for a fixed window, a meaningful anytime guarantee is not attainable, and a meaningful guarantee then requires a finite monitoring horizon.

Several extensions are of interest. First, abstentions are the price of guarantees without any assumption on the distributional shifts. Under assumptions that limit the size or frequency of the shifts, abstentions could be replaced by a wide but bounded set. Second, the learning rate $\gamma$ and the buffer $B$ in Algorithm \ref{alg:roll} are tuning parameters. Aggregating several values in the spirit of \citet{zaffran_2022} and \citet{gibbs_candes_2024} would remove the need to choose them a priori. Third, like PAC-type and anytime valid prediction sets, which are conservative even for IID data \citep{bates_2021, avelin_2023, scharfstein_2026, hultberg_2026}, our procedures are conservative: the true average miscoverage rate stays below $\alpha$ by about the margin $\omega_\tau$. How to reduce this conservativeness, for example, with tighter confidence sequences or assumptions on the shifts, is left for future research.

\spacingset{1}
\bibliographystyle{Chicago}
\bibliography{bibfile}

\newpage

\setcounter{page}{1}
\thispagestyle{plain}
\pagenumbering{Roman}

\appendix
\counterwithin{figure}{section}
\counterwithin{table}{section}
\counterwithin{equation}{section}
\spacingset{1.49}

\if0\blind { \begin{center}
  {\LARGE \bf Online Appendix for\\[1ex] ``Anytime Valid Conformal Prediction on Streaming Time Series''}

  \bigskip

  \renewcommand*{\thefootnote}{\fnsymbol{footnote}}

  {\large Chencheng Fang\footnotemark[1]\footnotemark[2]
  \footnotetext[1]{Institute of Finance and Statistics, University of Bonn, Adenauerallee 24-42, 53113 Bonn, Germany}
  \footnotetext[2]{Hausdorff Center for Mathematics, Endenicher Allee 62, 53115 Bonn, Germany}
  }

  \renewcommand*{\thefootnote}{\arabic{footnote}}
  \end{center}
} \fi


This online appendix is a supplement to the paper entitled ``Anytime Valid Conformal Prediction on Streaming Time Series''. Section \ref{app:derivations} contains our derivations; Section \ref{app:proofs} contains proofs; and Section \ref{app:noshift} contains some additional simulation results.

\section{Derivations}\label{app:derivations}

\subsection{\texorpdfstring{Derivation of Inequality \eqref{eq:feasible}}{Derivation of the Feasibility Inequality}}\label{app:feasible.ineq}

Consider a procedure that satisfies \eqref{eq:running_goal} under every data-generating process. Let $P$ be any underlying data-generating process and $t^*\ge1$. Fix $K>0$ and $\varepsilon\in(0,1)$, and let $P^*$ be the shifted process used in the proof of Proposition \ref{prop:no.pac}: $(\mathcal{D}_{t^*-1},X_{t^*})$ has the same distribution as under $P$, and $Y_{t^*}\sim N(\mu^*,1)$ independently of $(\mathcal{D}_{t^*-1},X_{t^*})$, where $\mu^*$ is so large that $\mathbb{P}(|Y_{t^*}|\le K)\le\varepsilon$.

Under $P^*$, if $\mathcal{C}_{t^*-1}(X_{t^*})\subseteq[-K,K]$, then $\nu_{t^*}\ge\mathbb{P}(|Y_{t^*}|>K)\ge1-\varepsilon$, so that $t^*\bar\nu_{t^*}\ge\sum_{t<t^*}\nu_t+1-\varepsilon$. If, in addition, $\sum_{t<t^*}\nu_t+1-\varepsilon>\alpha t^*$, then $\bar\nu_{t^*}>\alpha$. Since the procedure satisfies \eqref{eq:running_goal} under $P^*$, this has probability at most $\delta$:
\begin{equation*}
  \mathbb{P}\Big(\mathcal{C}_{t^*-1}(X_{t^*})\subseteq[-K,K]\ \text{ and }\ \sum_{t<t^*}\nu_t+1-\varepsilon>\alpha t^*\Big)\le\delta .
\end{equation*}
This event is determined by $(\mathcal{D}_{t^*-1},X_{t^*})$, whose distribution is the same under $P$ and $P^*$, so the inequality also holds under $P$. Letting $K\to\infty$ and $\varepsilon\to0$, the event increases to the event that $\mathcal{C}_{t^*-1}(X_{t^*})$ is bounded and $\sum_{t<t^*}\nu_t+1>\alpha t^*$, which therefore has probability at most $\delta$ under $P$. Taking the complement gives \eqref{eq:feasible}. If $t^*<1/\alpha$, then $\sum_{t<t^*}\nu_t+1\ge1>\alpha t^*$, so that $\mathcal{C}_{t^*-1}(X_{t^*})$ is bounded with probability at most $\delta$.

\subsection{\texorpdfstring{Derivation of Margins \eqref{eq:margin.stitched}--\eqref{eq:margin.normal}}{Derivation of Margins}}\label{app:margins}

By the proof of Lemma \ref{lem:skip}, the sum $\sum_{t\le\tau}(\nu_t-Z_t)=\tau(\bar\nu_\tau-\bar Z_\tau)$ is a martingale whose increments lie in $[-1,1]$ and are zero at abstentions. Each scaled margin $\tau\omega_\tau$ is an upper bound for this martingale that holds for all $\tau\ge1$ with probability at least $1-\delta$. Recall that $N_\tau$ is the number of bounded sets up to time $\tau$, and that $\widehat V_\tau$ sums $(Z_t-S_{t-1}/N_{t-1})^2$ over the steps $t\le\tau$ with a bounded set.

\begin{enumerate}[label={(\arabic*)}, align=left, leftmargin=*]
  \setlength\itemsep{-0.2em}
  \item Stitched Empirical Bernstein Margin: 
  
  We apply \citet[Theorem 4]{howard_2021} to the $[0,1]$-valued misses $Z_t$, whose conditional means $\nu_t$ are arbitrary, with the predictable sequence that equals $S_{t-1}/N_{t-1}$ at the steps with a bounded set and zero at abstentions. An abstention then adds nothing to the martingale or to $\widehat V_\tau$. By this theorem, the martingale is sub-exponential with variance process $\widehat V_\tau$ and scale one, and their stitched boundary (their Eq.\ (3.3) with $\eta=2$, $s=1.4$ and $m=1$ in their notation), evaluated at $\widehat V_\tau\vee1$, is crossed with probability at most $\delta$. With the constants rounded up, this boundary is the scaled margin in \eqref{eq:margin.stitched}. In \eqref{eq:margin.stitched}, we replace $\log\log(2(\widehat V_\tau\vee1))$ by its positive part: this only enlarges the margin, so that the bound remains valid, and it makes this term zero at the first bounded set, which simplifies the constant $c$ (Appendix \ref{app:c}).

  \item Hoeffding Margin: 
  
  At a step with a bounded set, $Z_t\in[0,1]$ has conditional mean $\nu_t$, so that Hoeffding's lemma gives $\mathbb{E}[\exp\{\lambda(\nu_t-Z_t)\}\mid\mathcal{F}_{t-1}]\le\exp(\lambda^2/8)$ for all $\lambda\ge0$, whatever the value of $\nu_t$. The martingale is therefore sub-Gaussian with variance process $N_\tau/4$. The same stitched boundary \citep[Theorem 1]{howard_2021}, evaluated at $N_\tau/4\vee1$ and without the linear term, which is not needed for sub-Gaussian processes, gives \eqref{eq:margin.hoeffding}.

  \item Normal Mixture Margin: 
  
  For the same sub-Gaussian martingale with variance process $N_\tau/4$, the one-sided normal mixture boundary of \citet[Proposition S2]{howard_2021} with tuning constant $\rho>0$ is the scaled margin in \eqref{eq:margin.normal}, and it is crossed with probability at most $\delta$.
\end{enumerate}

The same three bounds also hold for $\sum_{t\le\tau}(Z_t-\nu_t)$. Using each of the two one-sided bounds at error level $\delta/2$ gives, by a union bound, the two-sided margin $\omega'_\tau$ in \eqref{eq:cs.two}.

\subsection{\texorpdfstring{Derivation of the Constant $c$ in Section \ref{ssec:running}}{Derivation of the Constant c}}\label{app:c}

The constant $c$ bounds the one-step increase of the certified sum, $S^c_t-S^c_{t-1}=Z_t+\{t\omega_t-(t-1)\omega_{t-1}\}$, at every step $t$. Since $Z_t\le1$, it suffices to show that the scaled margin $t\omega_t$ increases by at most $c-1$ at every step. At an abstention, it does not change. At a step with a bounded set, we write the scaled margin as $g(v)$, where $v=\widehat V_t\vee1$ for \eqref{eq:margin.stitched}, $v=N_t/4\vee1$ for \eqref{eq:margin.hoeffding} and $v=N_t/4+\rho$ for \eqref{eq:margin.normal}. 

We proceed in two steps. First, at the first bounded set, $v=1$, $1$ and $1/4+\rho$, respectively, so that the scaled margin increases from zero to $g(1)$, $g(1)$ and $g(1/4+\rho)$. Setting $c-1$ equal to these values gives the three values of $c$ in Section \ref{ssec:running}. Second, at a later bounded set, $v$ increases by at most one for \eqref{eq:margin.stitched}, since $(Z_t-S_{t-1}/N_{t-1})^2\le1$, and by at most $1/4$ for \eqref{eq:margin.hoeffding} and \eqref{eq:margin.normal}. We show below, for each margin, that the resulting increase of $g$ is at most its increase at the first bounded set, so that the same $c$ also bounds the later steps.

\begin{enumerate}[label={(\arabic*)}, align=left, leftmargin=*]
  \setlength\itemsep{-0.2em}
  \item Stitched Empirical Bernstein Margin: 
  
  Let $a=0.72\log(5.19/\delta)$ and $\ell(v)=(\log\log(2v))_+$, so that $g(v)=1.7\sqrt{v\{\ell(v)+a\}}+3.4\,\ell(v)+2.45\log(5.19/\delta)$. For every $\delta\in(0,1)$, $a>1$. The function $\ell$ is continuous with $\ell(1)=0$, and its slope $\ell'(v)$ is at most $1/v\le1$ for $v\ge1$. Hence $\ell(v)\le v-1$ and $\ell(v)+a\le av$. The slope of $\sqrt{v\{\ell(v)+a\}}$ is therefore
\begin{equation*}
  \frac{\ell(v)+a+v\ell'(v)}{2\sqrt{v\{\ell(v)+a\}}}\le\frac{\ell(v)+a+1}{2\sqrt{v\{\ell(v)+a\}}}\le\sqrt{\frac{\ell(v)+a}{v}}\le\sqrt a,
\end{equation*}
where the second inequality uses $1<a\le\ell(v)+a$. Since the slope of $3.4\,\ell(v)$ is at most $3.4$, the slope of $g$ is at most $1.7\sqrt a+3.4\le1.7\sqrt a+2.45\log(5.19/\delta)=g(1)$. Since $v$ increases by at most one, $g$ increases by at most $g(1)$.

  \item Hoeffding Margin: 
  
  Here $g(v)=1.7\sqrt{v\{\ell(v)+a\}}$, with $a$ and $\ell$ as in (1). By the display in (1), the slope of $g$ is at most $1.7\sqrt a=g(1)$. Since $v$ increases by at most $1/4$, $g$ increases by at most $g(1)/4\le g(1)$.

  \item Normal Mixture Margin: 
  
  Here $g(v)=\sqrt{2v\,\ell(v)}$, where $\ell(v)$ now denotes the logarithm in \eqref{eq:margin.normal}, $\ell(v)=\log\{(2\delta)^{-1}\sqrt{v/\rho}+1\}$. For $\delta\le1/2$ and $v\ge\rho$, $\ell(v)\ge\log\{(2\delta)^{-1}+1\}>1/2$, while $v\ell'(v)\le1/2$. The slope of $g$ is therefore
\begin{equation*}
  \frac{\ell(v)+v\ell'(v)}{\sqrt{2v\,\ell(v)}}\le\frac{2\ell(v)}{\sqrt{2v\,\ell(v)}}=\sqrt{\frac{2\ell(v)}{v}}=\frac{g(v)}{v},
\end{equation*}
and $g(v)/v$ is non-increasing in $v$, because $v\ell'(v)\le1/2<\ell(v)$. For $v\ge1/4+\rho$, the slope of $g$ is thus at most $g(1/4+\rho)/(1/4+\rho)$. Since $v$ increases by $1/4$, $g$ increases by at most $g(1/4+\rho)/(1+4\rho)\le g(1/4+\rho)$.
\end{enumerate}

\subsection{\texorpdfstring{Derivation of Inequality \eqref{eq:cs.roll}}{Derivation of the Rolling Confidence Bound}}\label{app:cs.roll}

Consider the window ending at time $\tau\ge1$, which contains the time points $s_\tau+1,\dots,\tau$. The sum of the true miscoverage rates over this window is the difference of two running sums:
\begin{equation*}
  k_\tau\bar\nu^{(k)}_\tau=\sum_{t=s_\tau+1}^{\tau}\nu_t=\tau\bar\nu_\tau-s_\tau\bar\nu_{s_\tau}.
\end{equation*}
By Part (2) of Lemma \ref{lem:skip}, with probability at least $1-\delta$, $|\bar\nu_t-\bar Z_t|\le\omega'_t$ for all $t\ge1$, and then, we have $\tau\bar\nu_\tau\le S_\tau+\tau\omega'_\tau$ and $s_\tau\bar\nu_{s_\tau}\ge S_{s_\tau}-s_\tau\omega'_{s_\tau}$ (trivially if $s_\tau=0$). Hence
\begin{equation*}
  k_\tau\bar\nu^{(k)}_\tau\le(S_\tau-S_{s_\tau})+\tau\omega'_\tau+s_\tau\omega'_{s_\tau}.
\end{equation*}
Dividing by $k_\tau$ and using $S_\tau-S_{s_\tau}=k_\tau\bar Z^{(k)}_\tau$ gives \eqref{eq:cs.roll}. The upper bound for the rolling average thus uses the upper bound for the running average at the end of the window and the lower bound at its start, which is why the two-sided confidence sequence in Part (2) of Lemma \ref{lem:skip} is needed.

\subsection{Derivation of the Feasibility Check in Remark \ref{rmk:feasible.roll}}\label{app:feasible}

The feasibility check enforces the following: a bounded set is issued at time $t$ only if, whatever the value of $Z_t$, every window that contains $t$ stays within its allowance even if all later steps abstain. For this, we use the certified sum \eqref{eq:certified.sum.roll} for windows that start at any $s<t$, not only at $s_t$:
\begin{equation*}
  C(s,t)=\min\{N_t-N_s,\ S_t-S_s+t\,\omega'_t+s\,\omega'_s\}.
\end{equation*}
\begin{enumerate}[label={(\arabic*)}, align=left, leftmargin=*]
  \setlength\itemsep{-0.2em}
  \item Windows to check: A window that contains $t$ ends at some time $u\ge t$ and starts at $s_u\le t-1$. Since the start of the window never moves backwards, $s_u\ge s_t$. The windows that contain $t$ therefore start at some $s\in\{s_t,\dots,t-1\}$. As in Remark \ref{rmk:feasible.roll}, $\tau_s=\min\{u:s_u=s\}$ is the first time at which a window starts at $s$, with $\tau_s=\infty$ if no window starts at $s$; in that case, there is nothing to check.

  \item The allowance: A window that contains $t$ and starts at $s$ ends at a time $u$ with $u\ge t$ and, by the definition of $\tau_s$, $u\ge\tau_s$. Its length is therefore $k_u=u-s\ge\max(t,\tau_s)-s=m_t(s)$, and its allowance $\alpha k_u$ is at least $\alpha\,m_t(s)$.

  \item The certified sum: When the procedure decides whether to issue a bounded set at $t$, $Z_t$ is not yet known. Both possible values $z\in\{0,1\}$ must therefore be considered: $C^{(z)}(s,t)$ is $C(s,t)$ computed with $N_t=N_{t-1}+1$, $S_t=S_{t-1}+z$ and the margin $\omega'^{(z)}_t$ that results from a bounded set at $t$ with $Z_t=z$, as in Algorithm \ref{alg:roll}. If a bounded set is issued, then $C(s,t)=C^{(Z_t)}(s,t)$. At an abstention, $N_t$, $S_t$ and the scaled margin $t\,\omega'_t$ do not change, so that $C(s,t)=C(s,t-1)$ for every $s<t-1$. If all steps after $t$ abstain, the certified sum of a window that contains $t$ and starts at $s$ therefore stays equal to $C^{(Z_t)}(s,t)$.
\end{enumerate}
Combining (1)--(3), the requirement holds if $C^{(z)}(s,t)\le\alpha\,m_t(s)$ for both $z\in\{0,1\}$ and every $s\in\{s_t,\dots,t-1\}$ with $\tau_s<\infty$, which is the check in Remark \ref{rmk:feasible.roll} and Algorithm \ref{alg:roll}. A window that passes the check at $t$ thus stays within its allowance until the next bounded set is issued, at which the check is repeated.

For the running average, $k_t=t$ and only $s=0$ is checked. The check is then $C^{(z)}(0,t)\le\alpha t$, i.e., the certified sum after a possible miss at $t$ must not exceed $\alpha t$, which is the data-driven version of \eqref{eq:feasibility_running}. Algorithm \ref{alg:run}, however, does not need to carry out this check: whenever $\eta_t>0$, the buffer $B\ge\alpha/\gamma-\alpha+c$ in \eqref{eq:running.new.rule} gives $S^c_{t-1}<\alpha t-c$, and the certified sum increases by at most $c$, so that $S^c_t<\alpha t$ whatever the value of $Z_t$ (proof of Theorem \ref{thm:run}). 

\section{Proofs}\label{app:proofs}

\subsection{Proof of Proposition \ref{prop:aci.conv}}

\begin{proof}
  Summing both sides of the updating rule in \eqref{eq:aci} up to time $\tau$ gives \eqref{eq:aci.rule.sum}, i.e.,
  \begin{equation*}
    \frac{1}{\tau}\sum_{t=1}^{\tau}Z_t=\alpha+\frac{\alpha-\alpha_{\tau+1}}{\gamma\tau}.
  \end{equation*}
  It therefore suffices to show that $\alpha_t>-\gamma$ for all $t\ge1$, which we do by induction. This holds for $t=1$, since $\alpha_1=\alpha>0$. Suppose that it holds for $t$. If $\alpha_t\le0$, the prediction set is the whole real line, $Z_t=0$, and $\alpha_{t+1}=\alpha_t+\gamma\alpha>-\gamma$ by the induction hypothesis. If $\alpha_t>0$, then $\alpha_{t+1}\ge\alpha_t-\gamma(1-\alpha)>-\gamma$.

  An induction of the same kind gives $\alpha_t<1+\gamma\alpha$ for all $t\ge1$, which is used in the proof of Proposition \ref{prop:aci.window}. This holds for $t=1$. Suppose that it holds for $t$. If $\alpha_t\ge1$, the prediction set is empty, $Z_t=1$, and $\alpha_{t+1}=\alpha_t-\gamma(1-\alpha)<1+\gamma\alpha$. If $\alpha_t<1$, then $\alpha_{t+1}\le\alpha_t+\gamma\alpha<1+\gamma\alpha$.
\end{proof}

\subsection{Proof of Proposition \ref{prop:no.pac}}

\begin{proof}
  Fix $\varepsilon\in(0,1)$. Since $\mathcal{C}_{t^*-1}(X_{t^*})$ is bounded with probability one under $P$, there is a $K>0$ such that $\mathcal{C}_{t^*-1}(X_{t^*})\subseteq[-K,K]$ with probability at least $1-\varepsilon$. Let $P^*$ be the shifted process under which $(\mathcal{D}_{t^*-1},X_{t^*})$ has the same distribution as under $P$, $Y_{t^*}\sim N(\mu^*,1)$ independently of $(\mathcal{D}_{t^*-1},X_{t^*})$, where $\mu^*$ is so large that $\mathbb{P}(|Y_{t^*}|\le K)\le\varepsilon$, and the later observations are arbitrary. The prediction set $\mathcal{C}_{t^*-1}(X_{t^*})$ is determined by $(\mathcal{D}_{t^*-1},X_{t^*})$, so that it has the same distribution under $P^*$ as under $P$. Under $P^*$, if $\mathcal{C}_{t^*-1}(X_{t^*})\subseteq[-K,K]$, then
  \begin{equation*}
    \nu_{t^*}=\mathbb{P}\big(Y_{t^*}\notin\mathcal{C}_{t^*-1}(X_{t^*})\mid\mathcal{F}_{t^*-1}\big)\ge\mathbb{P}(|Y_{t^*}|>K)\ge1-\varepsilon,
  \end{equation*}
  because $Y_{t^*}$ is independent of $\mathcal{F}_{t^*-1}$. Since $\mathcal{C}_{t^*-1}(X_{t^*})\subseteq[-K,K]$ with probability at least $1-\varepsilon$, $\mathbb{P}(\nu_{t^*}\ge1-\varepsilon)\ge1-\varepsilon$.
\end{proof}

\subsection{Proof of Lemma \ref{lem:skip}}

\begin{proof}
  By assumption, whether a bounded set is issued at time $t$ is determined by $\mathcal{F}_{t-1}$. By \eqref{eq:Z}, $\mathbb{E}[\nu_t-Z_t\mid\mathcal{F}_{t-1}]=0$ at every step, and at an abstention, $\nu_t=Z_t=0$. Hence
  \begin{equation*}
    \sum_{t\le\tau}(\nu_t-Z_t)=\tau(\bar\nu_\tau-\bar Z_\tau),\qquad\tau\ge1,
  \end{equation*}
  is a martingale with respect to $(\mathcal{F}_t)_{t\ge0}$, whatever the distributional shifts. Its increments lie in $[-1,1]$ and are zero at abstentions, so that only the steps with a bounded set matter. A one-sided confidence sequence at error level $\delta$ bounds this martingale from above by the scaled margin $\tau\omega_\tau$ for all $\tau\ge1$ with probability at least $1-\delta$; the three margins \eqref{eq:margin.stitched}--\eqref{eq:margin.normal} are derived in Appendix \ref{app:margins}. Dividing by $\tau$ gives Part (1). A two-sided confidence sequence at error level $\delta$ bounds the absolute value of the martingale by $\tau\omega'_\tau$, which gives Part (2).
\end{proof}

\subsection{Proof of Theorem \ref{thm:run}}

\begin{proof}
  \begin{enumerate}[label={(\arabic*)}, align=left, leftmargin=*]
    \setlength\itemsep{-0.2em}
    \item Whether a bounded set is issued at time $t$, i.e., whether $\eta_t>0$, is determined by $\mathcal{F}_{t-1}$. By Part (1) of Lemma \ref{lem:skip} and Part (2), which is proved below, with probability at least $1-\delta$, $\bar\nu_\tau\le\bar Z_\tau+\omega_\tau\le\alpha$ for all $\tau\ge1$.

    \item We show by induction that $S^c_t\le\alpha t$ for all $t\ge0$. This holds for $t=0$, since $S^c_0=0$. Suppose that $S^c_{t-1}\le\alpha(t-1)$. If $\eta_t\le0$, the procedure abstains, so that $S^c_t=S^c_{t-1}\le\alpha t$. If $\eta_t>0$, the updating rule \eqref{eq:running.new.rule} gives $S^c_{t-1}<\alpha(t-1)+\alpha/\gamma-B\le\alpha t-c$, since $B\ge\alpha/\gamma-\alpha+c$. The certified sum increases by at most $c$ (Appendix \ref{app:c}), so that $S^c_t\le S^c_{t-1}+c<\alpha t$. Dividing $S^c_\tau\le\alpha\tau$ by $\tau$ gives $\bar Z_\tau+\omega_\tau\le\alpha$, and $\bar Z_\tau\le\alpha$ because $\omega_\tau\ge0$.

    \item Since $\tau\omega_\tau$ is non-decreasing, $S^c_t-S^c_{t-1}\ge Z_t$ for all $t\ge1$. As in the proof of Proposition \ref{prop:aci.conv}, this gives $\eta_t<1+\gamma\alpha$ for all $t\ge1$ by induction. This holds for $t=1$, since $\eta_1=\alpha-\gamma B<1$. Suppose that it holds for $t$. If $\eta_t\ge1$, the prediction set is empty, $Z_t=1$, and \eqref{eq:running.new.rule.rec} gives $\eta_{t+1}\le\eta_t-\gamma(1-\alpha)<1+\gamma\alpha$. If $\eta_t<1$, then $\eta_{t+1}\le\eta_t+\gamma\alpha<1+\gamma\alpha$. Rearranging \eqref{eq:running.new.rule} and plugging in $\eta_{\tau+1}<1+\gamma\alpha$,
    \begin{equation*}
      S^c_\tau=\alpha\tau-B+\frac{\alpha-\eta_{\tau+1}}{\gamma}>\alpha\tau-B-\frac{1-\alpha}{\gamma}-\alpha .
    \end{equation*}
    The condition holds for the margins \eqref{eq:margin.stitched}--\eqref{eq:margin.normal}: the scaled margin is zero before the first bounded set, does not change at an abstention, and is a non-decreasing function of $\widehat V_\tau$ or $N_\tau$, neither of which decreases. \qedhere
  \end{enumerate}
\end{proof}

\subsection{Proof of Proposition \ref{prop:aci.window}}

\begin{proof}
  Summing both sides of the updating rule in \eqref{eq:aci} over the window gives \eqref{eq:aci.rule.window}, i.e., $S_\tau-S_{s_\tau}=\alpha k_\tau+(\alpha_{s_\tau+1}-\alpha_{\tau+1})/\gamma$. By the proof of Proposition \ref{prop:aci.conv}, $\alpha_{s_\tau+1}<1+\gamma\alpha$ and $\alpha_{\tau+1}>-\gamma$, so that $S_\tau-S_{s_\tau}<\alpha k_\tau+\{1+\gamma(1+\alpha)\}/\gamma$. Dividing by $k_\tau$ gives the claim.
\end{proof}

\subsection{Proof of Theorem \ref{thm:roll}}

\begin{proof}
  Unlike in the proof of Theorem \ref{thm:run}, the updating rule \eqref{eq:eta.roll} alone does not bound the certified sum, because $C(s_t,t)$ can increase at abstentions. The bound comes from the feasibility check instead. We use $C(s,t)$, $\tau_s$ and $m_t(s)$ as in Remark \ref{rmk:feasible.roll} and Appendix \ref{app:feasible}.

  \begin{enumerate}[label={(\arabic*)}, align=left, leftmargin=*]
    \setlength\itemsep{-0.2em}
    \item Whether a bounded set is issued at time $t$ is determined by $\mathcal{F}_{t-1}$, so that Lemma \ref{lem:skip} applies. By Appendix \ref{app:cs.roll}, with probability at least $1-\delta$, for all $\tau\ge1$,
    \begin{equation*}
      \sum_{t=s_\tau+1}^{\tau}\nu_t\le S_\tau-S_{s_\tau}+\tau\omega'_\tau+s_\tau\omega'_{s_\tau}.
    \end{equation*}
    The left-hand side is also at most $N_\tau-N_{s_\tau}$, the number of bounded sets in the window, since $\nu_t\le1$ at a step with a bounded set and $\nu_t=0$ at an abstention. It is therefore at most the smaller of the two bounds, which is $C(s_\tau,\tau)$ by \eqref{eq:certified.sum.roll}. By Part (2), which is proved below, $C(s_\tau,\tau)\le\alpha k_\tau$. Hence $\sum_{t=s_\tau+1}^{\tau}\nu_t\le\alpha k_\tau$, i.e., $\bar\nu^{(k)}_\tau\le\alpha$, for all $\tau\ge1$.

    \item Fix $\tau\ge1$ and write $s=s_\tau$ for the start of the window ending at $\tau$. We show that $C(s,\tau)\le\alpha k_\tau$. The idea is that, for this fixed start $s$, the certified sum $C(s,\cdot)$ changes only when a bounded set is issued, and every bounded set had to pass the feasibility check. We distinguish two cases.

    Case 1: no bounded set is issued at the steps $s+1,\dots,\tau$. Then $N_\tau-N_s=0$, so that $C(s,\tau)=0\le\alpha k_\tau$ by \eqref{eq:certified.sum.roll}.

    Case 2: at least one bounded set is issued at the steps $s+1,\dots,\tau$. Let $t$ be the last such step. Since the start of the window never moves backwards, $s_t\le s\le t-1$, and $\tau_s\le\tau$. The check at time $t$ therefore included the start $s$ and was passed for both $z\in\{0,1\}$. With $z=Z_t$,
    \begin{equation*}
      C(s,t)=C^{(Z_t)}(s,t)\le\alpha\,m_t(s)=\alpha\{\max(t,\tau_s)-s\}\le\alpha(\tau-s)=\alpha k_\tau .
    \end{equation*}
    The steps $t+1,\dots,\tau$ are abstentions, at which the certified sum does not change (Appendix \ref{app:feasible}), so that $C(s,\tau)=C(s,t)\le\alpha k_\tau$.

    In both cases, $C(s_\tau,\tau)\le\alpha k_\tau$. Finally, $S_\tau-S_{s_\tau}\le C(s_\tau,\tau)$, because both terms in \eqref{eq:certified.sum.roll} are at least $S_\tau-S_{s_\tau}$: misses occur only at steps with a bounded set, and the margins are non-negative. Hence $\bar Z^{(k)}_\tau\le\alpha$. \qedhere
  \end{enumerate}
\end{proof}

\subsection{Proof of Proposition \ref{prop:regimes}}

\begin{proof}
  We bound the difference $C(s_\tau,\tau)-(S_\tau-S_{s_\tau})$ between the certified and the realized sum of misses in the window by a deterministic function $h(\tau)$, and then compare $h(\tau)$ with $k_\tau$.

  First, by \eqref{eq:certified.sum.roll}, $C(s_\tau,\tau)-(S_\tau-S_{s_\tau})\le\tau\omega'_\tau+s_\tau\omega'_{s_\tau}$. As in Appendix \ref{app:c}, with $\delta/2$ in place of $\delta$, each scaled margin $t\,\omega'_t$ is zero or equals $g(v)$, where $g$ is non-decreasing and $v\le t$ for \eqref{eq:margin.stitched}--\eqref{eq:margin.hoeffding} and $v\le t+\rho$ for \eqref{eq:margin.normal}. With $h(\tau)=2g(\tau)$ for \eqref{eq:margin.stitched}--\eqref{eq:margin.hoeffding}, $h(\tau)=2g(\tau+\rho)$ for \eqref{eq:margin.normal}, and $s_\tau\le\tau$, both scaled margins are at most $h(\tau)/2$, so that $C(s_\tau,\tau)-(S_\tau-S_{s_\tau})\le h(\tau)$.

  Second, by \eqref{eq:margin.stitched}--\eqref{eq:margin.normal}, $g(v)$ grows at most like $\sqrt{v\log v}$ and at least like $\sqrt v$, so that $h(\tau)=O(\sqrt{\tau\log\tau})$ and $h(\tau)/\sqrt\tau$ is bounded away from zero.

  Finally, we divide by $k_\tau$. For proportional windows, $k_\tau\ge\theta\tau$, so that $h(\tau)/k_\tau=O(\sqrt{\log\tau/\tau})$, which is Part (1). For polynomial windows, $k_\tau\ge\tau^\beta$, so that $h(\tau)/k_\tau=O(\tau^{1/2-\beta}\sqrt{\log\tau})$, which is Part (2). For Part (3), write $h(\tau)/k_\tau=\{h(\tau)/\sqrt\tau\}\{\sqrt\tau/k_\tau\}$. The first factor is bounded away from zero by the second step, and so is the second factor when $k_\tau=O(\sqrt\tau)$. Hence $h(\tau)/k_\tau$ does not vanish.
\end{proof}

\section{Additional Simulation Results}\label{app:noshift}

\subsection{True Running Average Miscoverage Rate for Stable Scenario}

\begin{figure}[!htbp]
\centering
\maybeinclude[width=0.8\textwidth]{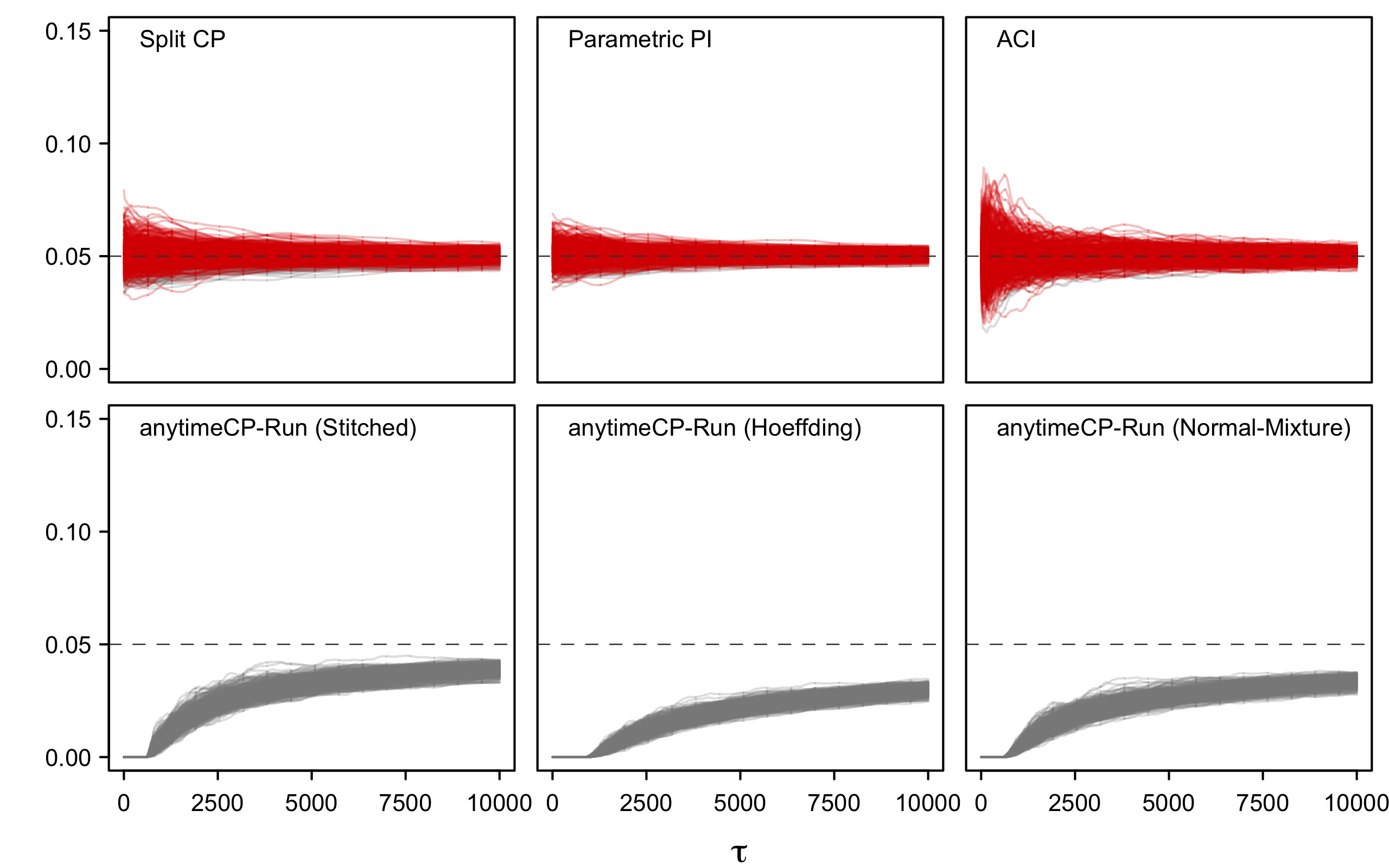}
\caption{True running average $\bar\nu_\tau$ of each replication without distributional shifts.}
\label{fig:sim.run.paths.noshift}
\end{figure}

\begin{table}[!htbp]
\centering\spacingset{1}\footnotesize
\setlength{\tabcolsep}{4pt}\renewcommand{\arraystretch}{1.0}
\begin{tabular}{lcccccccc}
\toprule
& \multicolumn{2}{c}{$\max_{\tau\ge 1}\bar\nu_\tau$} & \multicolumn{3}{c}{Abstention} & \multicolumn{3}{c}{Width} \\
\cmidrule(lr){2-3}\cmidrule(lr){4-6}\cmidrule(lr){7-9}
& median & $95\%$ & I & II & III & I & II & III \\
\midrule
Split CP & 0.054 & 0.065 & 0 & 0 & 0 & 3.93 & 3.93 & 3.94 \\
Parametric PI & 0.054 & 0.061 & 0 & 0 & 0 & 3.93 & 3.93 & 3.93 \\
ACI & 0.060 & 0.073 & 0 & 0 & 0 & 3.94 & 3.94 & 3.94 \\
anytimeCP-Run (Stitched) & 0.039 & 0.042 & 16\% & 0 & 0 & 4.22 & 4.07 & 4.04 \\
anytimeCP-Run (Hoeffding) & 0.030 & 0.033 & 23\% & 0 & 0 & 4.55 & 4.24 & 4.16 \\
anytimeCP-Run (Normal-Mixture) & 0.033 & 0.037 & 15\% & 0 & 0 & 4.41 & 4.18 & 4.12 \\
\bottomrule
\end{tabular}
\caption{Summary statistics of true running average miscoverage rate for all replications without distributional shifts.}
\label{tab:sim.run.noshift}
\end{table}

Without shifts, Split CP, Parametric PI and ACI stay close to $\alpha$ but exceed it at some time in most replications. anytimeCP-Run stays below $\alpha$ in every replication and no longer abstains after Phase I. In Phase III, its prediction sets are only $3\%$, $6\%$ and $5\%$ wider than those of ACI.

\subsection{True Rolling Average Miscoverage Rate for Stable Scenario}

\begin{figure}[!htbp]
\centering
\maybeinclude[width=0.85\textwidth]{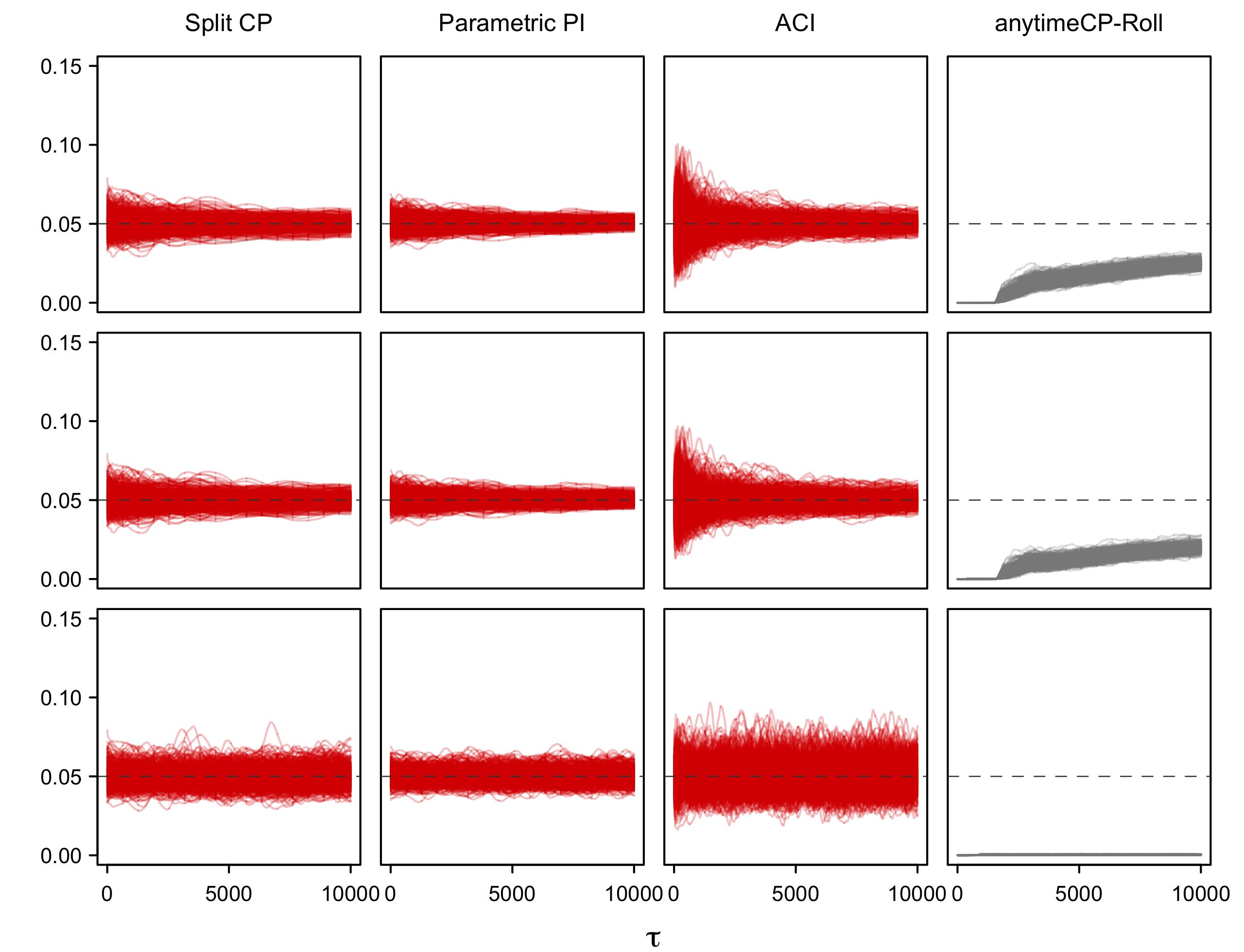}
\caption{True rolling average $\bar\nu^{(k)}_\tau$ of each replication without distributional shifts.}
\label{fig:sim.roll.paths.noshift}
\end{figure}

\begin{table}[!htbp]
\centering\spacingset{1}\footnotesize
\setlength{\tabcolsep}{4pt}\renewcommand{\arraystretch}{1.0}
\begin{tabular}{llcccccccc}
\toprule
& & \multicolumn{2}{c}{$\max_{\tau\ge1}\bar\nu^{(k)}_\tau$} & \multicolumn{3}{c}{Abstention} & \multicolumn{3}{c}{Width} \\
\cmidrule(lr){3-4}\cmidrule(lr){5-7}\cmidrule(lr){8-10}
Window & Method & median & $95\%$ & I & II & III & I & II & III \\
\midrule
proportional & Split CP & 0.057 & 0.067 & 0 & 0 & 0 & 3.93 & 3.93 & 3.94 \\
 & Parametric PI & 0.056 & 0.063 & 0 & 0 & 0 & 3.93 & 3.93 & 3.93 \\
 & ACI & 0.067 & 0.084 & 0 & 0 & 0 & 3.94 & 3.94 & 3.94 \\
 & anytimeCP-Roll (Prop) & 0.025 & 0.029 & 47\% & 3\% & 1\% & 4.91 & 4.65 & 4.49 \\
\midrule
polynomial & Split CP & 0.058 & 0.067 & 0 & 0 & 0 & 3.93 & 3.93 & 3.94 \\
 & Parametric PI & 0.056 & 0.063 & 0 & 0 & 0 & 3.93 & 3.93 & 3.93 \\
 & ACI & 0.066 & 0.082 & 0 & 0 & 0 & 3.94 & 3.94 & 3.94 \\
 & anytimeCP-Roll (Poly) & 0.020 & 0.024 & 52\% & 6\% & 3\% & 5.08 & 4.84 & 4.67 \\
\midrule
fixed & Split CP & 0.062 & 0.070 & 0 & 0 & 0 & 3.93 & 3.93 & 3.94 \\
 & Parametric PI & 0.060 & 0.065 & 0 & 0 & 0 & 3.93 & 3.93 & 3.93 \\
 & ACI & 0.073 & 0.085 & 0 & 0 & 0 & 3.94 & 3.94 & 3.94 \\
 & anytimeCP-Roll (Fixed) & 0.001 & 0.001 & 96\% & 96\% & 96\% & 5.21 & 5.19 & 5.19 \\
\bottomrule
\end{tabular}
\caption{Summary statistics of true rolling average miscoverage rate for all replications without distributional shifts.}
\label{tab:sim.roll.noshift}
\end{table}

Without shifts, anytimeCP-Roll stays below $\alpha$ in every replication. In Phase III, it abstains at $1\%$ and $3\%$ of the steps for proportional and polynomial windows, with prediction sets only $14\%$ and $19\%$ wider than ACI. For fixed windows, it remains valid but abstains at 96\% of the steps.

\end{document}